\documentstyle{l-aa}              % Option2 : Final A+A version (2 columns)
\input epsfig.sty

\def\mathrm{\rm}                   % please use this if your latex
\def\mathbf{\bf}                   % does not like \mathrm and \mathbf

\def\simless{\mathbin{\lower 1pt\hbox
   {$\spose{\raise 5pt\hbox{$\char'074$}}\char'430$}}}
\def\simgreat{\mathbin{\lower 1pt\hbox
   {$\spose{\raise 5pt\hbox{$\char'076$}}\char'430$}}}
\def\lapp{\mathbin{\raise2pt \hbox{$<$} \hskip-9pt \lower4pt \hbox{$\sim$}}}
\def\gapp{\mathbin{\raise2pt \hbox{$>$} \hskip-9pt \lower4pt \hbox{$\sim$}}}
\begin{document}

   \thesaurus{07         % A&A Section 7: Stellar Atmospheres.
              (02.13.2;  % MHD,
               06.19.2;  % Solar wind
               2.16.1;  % Stars: atmospheres
               08.23.3;  % Stars: winds, outflows
               08.16.5;  % Stars: pre-main sequence 
%              11.10.1;  % Galaxies: jets
               09.10.1)} % ISM: jets and outflows
   \title{Magnetic collimation of the solar and stellar winds} 

    \author{K. Tsinganos %$^{\dag}$\thanks{tsingan@physics.uch.gr}
           \inst{1}
   \and    S. Bogovalov %$^{\dag}$\thanks{bogoval@photon.mephi.ru}
           \inst{2}
          }
 
   \offprints{K. Tsinganos\protect,\\ tsingan@physics.uch.gr}
 
   \institute
         {Department of Physics, University of Crete and FORTH/IESL  
         GR-710 03 Heraklion, Crete, Greece
    \and Moscow State Engineering Physics Institute, Moscow, 115409, Russia
         }
 
    \date{Submitted July 21, 1999; accepted  February 3, 2000}
 
   \maketitle
   \markboth{K. Tsinganos \& S. Bogovalov:   
             Magnetic collimation of the solar and stellar winds}{}
   \begin{abstract}

We resolve the paradox that although magnetic collimation of an isotropic 
solar wind results in an enhancement of its proton flux along the polar 
directions, several observations indicate a wind proton flux peaked at the 
equator. 
To that goal, we solve the full set of the time-dependent MHD equations 
describing the axisymmetric outflow of plasma from the magnetized 
and rotating Sun, either in its present form of the solar wind, or, in its
earlier form of a protosolar wind.
Special attention is directed towards the collimation properties of the
solar outflow at large heliocentric distances.
For the present day solar wind it is found that the poloidal streamlines 
and fieldlines are only slightly focused toward the solar poles. 
However, even such a modest compression of the flow by the azimuthal 
magnetic field would lead to an increase of the mass
flux at the polar axis by about 20\% at 1 AU, relatively to its value at
the equator, for an initially isotropic at the base wind, contrary to 
older and recent (Prognoz, Ulysses, SOHO) observations.  
For the anisotropic in heliolatitude wind with parameters at the base 
inferred from {\it in situ} observations by ULYSSES/SWOOPS and SOHO/CDS 
the effect of collimation is almost totally compensated by the initial 
velocity and density anisotropy of the wind. 
This effect should be taken into account in the interpretation of the 
recent SOHO observations by the SWAN instrument.
Similar simulations have been performed for a five- and ten-fold increase
of the solar angular velocity corresponding presumably to the 
wind of an earlier phase of our Sun. 
For such conditions it is found that for initially radial streamlines,  
the azimuthal magnetic field created by the fast rotation focus them 
toward the rotation axis and forms a tightly collimated jet.

\vskip 0.5 true cm

   \end{abstract}

\section{Introduction}

Several stellar and extragalactic astrophysical systems have been observed 
to exhibit collimated outflows in the form of jets (young stellar objects, 
low and high mass X-ray binaries, black hole X-ray transients, symbiotic 
stars, planetary nebulae nuclei, supersoft X-ray sources, active galactic 
nuclei and quasars). In recent reviews of observations from all such 
classes of astrophysical objects it has been argued that an interconnecting 
element may be a rotating accretion disk threaded 
by a magnetic field (Livio 1999, K\"{o}nigl \& Pudritz, 1999). 
Such connection between the disk and the jet 
is particularly evident in HST observations of several young stars in the 
nearby Orion nebula (Ray 1996). This rather convincing observational evidence 
of a close jet-disk relation is the basis for the presently prevailing view 
that an accretion disk is the necessary ingredient for the production of 
collimated jets.

On the other hand, theoretically it has been shown for quite some time by 
now that gas outflows from a rotating magnetized object of any nature  
can be magnetically  self-collimated to form a jet (Heyvaerts \& Norman 
1989, Chiueh et al 1991, Sauty \& Tsinganos 1994, Bogovalov 1995).  
This result seems to be a rather intrinsic property of magnetized winds with 
polytropic  thermodynamics or not, where the self-compression of the plasma 
is provided by the toroidal magnetic field induced by the rotation of the 
central source. 
Henceforth emerges the generally accepted opinion that all observed jets 
are magnetically collimated (Livio 1999).
However, no direct observational evidence exists today that most observed 
jets are indeed collimated solely by magnetic fields. 
Recently, it has been pointed out that the toroidal magnetic field is 
unstable and cannot collimate the jet effectively (Spruit et al. 1997, 
Lucek \& Bell 1997) and it has been argued that 
magnetized winds do not collimate without an external help, such as 
the channelling effects of a thick accretion disk and/or confinement 
from the ambient medium (Okamoto 1999). It is thus crucially 
important to find direct observational evidence that the magnetic field 
mainly collimates the plasma in observed jets and by this 
way to test the theory of magnetic collimation.

Nevertheless, plasma outflows do also emanate from isolated magnetized 
and rotating stars without an accretion disk, of which the solar wind (SW) 
is the classical and best studied example.  The natural question which 
arises then is to what observable degree dynamical effects are 
capable to collimate outflows from such single stars too. 
Theoretical studies on the angular momentum evolution of solar-type 
stars have concluded that at the end of the early accretion phase (PTTS) 
the star may be span up by more than 10 times the present solar rotation rate 
while its magnetic field is also strong (Bouvier et al. 1997). 
And, in recent studies it has been shown that cold winds from such rapidly 
rotating and highly magnetized stars lead to considerable collimation 
of the outflow (Bogovalov \& Tsinganos 1999, henceforth Paper I). 
Similar is the result from studies of hot plasma outflows from efficient 
magnetic rotators (Sauty \& Tsinganos 1994, Sauty et al. 1999). 
Hence, observation of the collimation effect in outflows from single stars 
could be the most reliable observational test of the theory of magnetic 
collimation. 

The question of the degree of collimation of the SW is an interesting 
possibility that has not been fully answered theoretically 
and observationally for quite some time now. 
Suess (1972) and Nerney \& Suess (1975) were the first to model the 
axisymmetric interaction of magnetic fields with rotation in stellar 
winds by a linearisation of the MHD equations in inverse Rossby numbers 
and to find a poleward deflection of the streamlines of the solar wind 
caused by the toroidal magnetic field. Later, Sakurai (1985) addressed 
the same problem by numerically solving the system of the polytropic MHD 
equations for the stationary outflow. Washimi \& Shibata (1993) 
modelled time dependent axisymmetric thermo-centrifugal winds with a 
{\it dipole} magnetic flux distribution on the stellar surface and a 
radial field in Washimi \& Sakurai (1993). Polytropic MHD simulations 
of magnetized winds containing both a "wind" and a "dead" zone (Tsinganos 
\& Low 1989, Mestel 1999) 
have also been performed up to distances of 0.25 AU (Keppens \& Goedbloed 1999). 
All these studies show some small deflection of the flow toward the axis 
of rotation. 

In the observational front, information on the degree of collimation 
of the SW can be inferred from anisotropies in the Lyman alpha emission. 
These solar UV photons are scattered by neutral H atoms of interstellar 
origin and where the SW mass flux is increased the neutral H atoms are 
destroyed and thus the Lyman alpha emission is reduced. 
Early observations by the Mariner 10 (Kumar \& Broadfoot 1979) and 
Prognoz (Bertaux et al. 1985) 
satellites have shown that there is less Lyman alpha emission near the 
equator in comparison to the ecliptic poles, than predicted by an 
isotropic SW (Bertaux et al. 1997). Therefore, these 
Lyman alpha observations imply that the SW mass efflux should be 
maximum at the equator and minimum at the poles. The same trend is confirmed 
by {\it in situ} observations of Ulysses (Goldstein et al. 1996) and the 
SWAN instrument onboard of the SOHO spacecraft (Kyr\"{o}l\"{a} et al. 1998). 
However, the effect of SW collimation around the ecliptic poles would 
cause the opposite effect on Lyman alpha observations. In other words, 
although UV observations infer a SW mass efflux peaked at the equator, 
magnetic collimation would cause a SW mass efflux peaked at the poles, 
for an isotropic at the base wind. 
   
One of the main purposes of this paper is to resolve this paradox. 
We shall follow the idea of magnetic collimation of the SW and show  
which values of the parameters characterizing 
the heliolatitudinal dependance of the SW (Lima et al. 1997, Gallagher et al. 
1999), such as 
density, bulk flow speed, mass efflux, etc are consistent with the 
observations by the Lyman alpha method.  Furthermore, we shall follow 
the increase of the degree of collimation of a hot stellar wind by 
increasing the rotation rate of the star and show that a ten-fold 
increase of angular velocity, as is the case in the majority of the 
young rapid rotators, leads to a dramatic increase of the degree of 
stellar wind collimation.  

The paper is organised as follows. In Sect. 2 we justify the use of a 
split monopole model in our analysis for the collimation properties of the 
realistic solar wind at large distances from the Sun. In order to establish 
notation in Sect. 3 we give briefly the basic equations describing a 
stationary and polytropic Parker wind. In Sect. 4, the initial configuration
used together with the boundary conditions for the numerical simulation in the 
nearest zone are discussed. In Sect. 5 the analytical method for extending the 
integration to unlimited large distances outside the near zone is briefly 
described while in Sect. 6 the parametrization of the presented solutions 
is given.    
In Sect. 7 we discuss the results for the isotropic SW in the near zone 
containing the critical surfaces and in the asymptotic regime of the 
collimated outflow, for a uniform rotation.
In Sect. 8 the case of a stellar wind from a star rotating faster than the 
Sun is taken up. 
A brief summary with a discussion of the main results is finally given 
in Sect. 9.  
  
\section{The assumption of a split-monopole model for a stellar wind} 
%Figure 1
\begin{figure}
\vspace{-2.0cm}
\centerline{\psfig{file=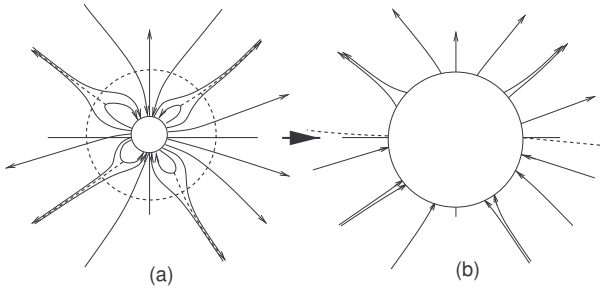,height=7.0truecm,angle=360}\hspace{2.0cm}}
\caption{Schematic drawing of a white light image of the solar 
atmosphere indicating the shape of the solar magnetic fieldlines. 
Beyond the dotted spherical distance located around the slow magnetosonic 
critical surface the magnetic fieldlines are radial to a first approximation. 
Electric sheet currents are located on the dotted lines.}
\label{schematic}
\end{figure}

Fig. 1a is a sketch of the magnetic field structure in the corona of a 
star. Close to the stellar base the structure of the magnetosphere 
may be rather complicated. In this paper we are interested in the study
of winds at distances much larger than the radius of the star  
where no closed field lines exist. It is therefore reasonable
to consider the plasma flow starting at distances shown in 
Fig. 1a  by the dashed line. For the solar wind the location of this 
surface can be put somewhere between the slow magnetosonic
surface and the Alfv\'enic surface. We choose this location of the   
starting surface to avoid the solution of the problem of the
wind acceleration in the very vicinity of the star which is 
defined not only by thermal pressure gradients but also by  
nonthermal processes of acceleration where the acceleration mechanisms 
have not still studied sufficienty well and are beyond the scope of the 
present study (e.g., see Holzer \& Leer 1997, Hansteen et al. 1997, 
Wang et al. 1998).  Above this base surface
we can assume that the dynamics of the wind is mainly controlled by 
thermal and electromagnetic forces. 
In this approach the density and velocity of the plasma, together 
with the tangential components of the electric field and the normal 
component of the magnetic field are specified on this base surface 
while the tangential components of the magnetic field are free.

%Figure 2
\begin{figure}
\centerline{\psfig{file=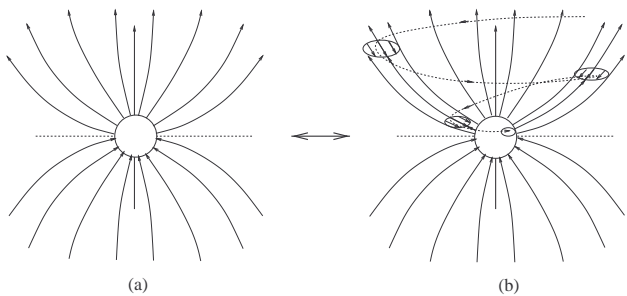,height=4.0truecm,angle=270}}
\caption{Plasma flow from an axisymmetric rotator with an initially split-monopole 
magnetic field, as in panel (a). Reversing the direction of the poloidal 
magnetic field in an arbitrary flux tube does not change the dynamics of 
the problem while we obtain the configuration shown in panel (b) which 
describes a nonstationary and nonaxisymmetric plasma flow from a rotator 
with a magnetic spot of opposite polarity on the
base surface. The distribution of such spots can be arbitrary. 
}
\label{schematic2}
\end{figure}

Nevertheless, the solution of this problem is still too complicated. 
Open poloidal magnetic field lines go to infinity and change their direction 
on the so called current sheets, some of which are indicated with dotted 
lines in Fig. 1a. These  current sheets are present in any realistic wind 
from a stellar atmosphere, since the total magnetic flux of the open poloidal 
magnetic field lines is equal to zero while the mass loss rate is finite. 
The invariance principle summarized in the Appendix in a form appropriate 
to hot winds in a gravitational field allows us to simplify the structure of
the magnetic field in the wind (Bogovalov 1999). 
According to this principle, we can
reverse the direction of some field lines so that the magnetic 
field is unipolar in each hemisphere, e.g., outward in the upper
and inwards in the lower hemisphere. In ideal MHD wherein we neglect all
dissipative processes such as magnetic reconnection, this operation does not affect
the dynamics of the plasma as long as the streamlines are not modified. 
This results in the configuration
shown in Fig. 1b where since we are not interested in the region 
upstream of the base surface, this region is not shown. 
In this structure the current sheet is located only around the equatorial plane.
To proceed, we further assume that the distribution of the
normal component of the poloidal magnetic field is uniform in the upper 
and lower hemispheres. In that case we get the model of the axisymmetric 
rotator with an initially split-monopole magnetic field. The field 
lines are magnetically focused toward the system's axis, as shown in 
Fig. 2a.  
%The flow can be considered in this case as stationary and axially symmetric.

We would like to stress that the model of the axisymmetric 
rotator describes not only axisymmetric outflows but also a wide class
of nonstationary and axially nonsymmetric outflows. 
This is due to the fact that according to the invariance principle  
(c.f. Appendix) the change of the direction of a magnetic field 
line in some flux tube does not affect the dynamics of the plasma. 
For example, let's assume that we obtain a solution for the axisymmetric 
rotator, as shown schematically in Fig. 2a.  
Then, a reversal of the sign of some magnetic field lines in an arbitrary 
poloidal flux tube gives us a solution which is not
axisymmetric and nonstationary, as shown in Fig. 2b. 
This is a solution for the plasma outflow from a rotator with uniform 
magnetic field at the base surface but with a magnetic spot 
of the opposite polarity on the upper hemisphere. 
Fig. 2b shows the cross-section of such a magnetic field by the 
poloidal plane. 
The stream lines are the same as for the axisymmetric case. 
But the poloidal magnetic field changes sign in magnetic spots 
corresponding to the flux tube of the opposite polarity. 
The path of the field line in this flux tube in 3D is shown by a 
dashed line. These spots propagate in the poloidal plane with the 
velocity of the plasma and hence the pattern is nonstationary.

It is clear that the number of such magnetic spots and their position 
at the base surface can be arbitrary. 
Therefore the study of the plasma outflow in the model of the axisymmetric 
rotator with an initially split-monopole magnetic field allows us 
to study a much more wider classes of nonstationary and nonaxisymmetric 
flows.

\section{The stationary polytropic Parker wind}

A Parker wind is taken as the initial state (t=0) for the solution of the
time-dependent problem. In this initial state, the wind is assumed to
flow along the radial magnetic field lines of an isotropic 
magnetic field (Parker 1963), although in general, the flow is not
isotropic such that the wind has its own integrals of motion on every
stream line $\psi=const$. For simplicity, a polytropic relationship
between the pressure $P$ and the density $\rho$ is assumed 
\begin{equation}
P = Q(\psi) \rho^\gamma
\,,
\end{equation}
where $\gamma$ is the polytropic index. Then, the Bernoulli equation
for energy conservation along a radial line in such an anisotropic 
wind from a nonrotating star has the form
\begin{equation}
{V^{2}\over 2}+{\gamma \over \gamma -1}Q(\psi) \rho^{\gamma -1} -
{GM\over R}=E(\psi)
\,.
\label{ber1}
\end{equation}

Denote by $V_{\infty}(\psi)$ the terminal velocity of the plasma on each fieldline.
In order to get equations in dimensionless variables, we shall
use for the radial distance $\tilde R=R/R_{\rm s, eq}$, the density
$\tilde\rho=\rho/\rho_{\rm s, eq}$, the entropy function $\tilde Q = Q/Q_{\rm eq}$
and the velocity $v=V/V_{\rm s, eq}$, in terms of the equatorial values
of the sonic distance $R_{\rm s, eq}$, density
$\rho_{\rm s, eq}$, entropy function $Q_{\rm eq}$ and sound speed $V_{\rm s, eq}$.

The mass flux conservation in these dimensionless variables takes the form
\begin{equation}
\tilde \rho v \tilde R^{2}  ={\dot m(\psi)}
%={\dot m(\psi)\over \dot m_{\rm eq}(\psi)}
\,,
\label{mass}
\end{equation}
while the Bernoulli integral becomes,
\begin{equation}
{v^{2}\over 2} + {\gamma \over \gamma -1} \tilde Q(\psi) 
\tilde\rho^{(\gamma - 1)}
-\left({GM\over V^{2}_{\rm s, eq}R_{\rm s, eq}}\right){1\over\tilde R}=
{v_{\infty}^{2}(\psi ) \over 2}
\,.
\label{ber2}
\end{equation}
Since $v_{\infty}(\psi )$ can be regarded as a function of $\tilde R$
and $\tilde \rho$ along a particular streamline $\psi$, by taking the partial 
derivative of $v_{\infty}^{2}$ with respect to $\tilde \rho$ and $\tilde R$ 
we obtain 
%\begin{equation}
%{\partial \over \partial \tilde \rho}
%v^2_{\infty} (\tilde \rho, \tilde R; \psi =const.)=
%{\partial \over \partial \tilde R} v^2_{\infty} (\tilde \rho, \tilde R; \psi =const.)= 0
% \,.
%\label{bernn}
%\end{equation}
%The two resulting 
the usual Parker criticality relations which give the sound speed $v_{\rm s}(\psi)$ 
and the spherical distance $\tilde R_{\rm s} (\psi)$ of the sonic critical surface 
along each streamline $\psi$ =const., 
\begin{equation}
\label{crit1}
v_{\rm s}^{2} (\psi) =\gamma \tilde Q(\psi)\tilde\rho_{\rm s}^{(\gamma-1)}(\psi)
\,,
\label{}
\end{equation}
and
\begin{equation}
\tilde R_{\rm s}(\psi) =
{GM\over 2 V^{2}_{\rm s, eq}R_{\rm s, eq}} {1\over v_{\rm s}^{2}(\psi) }
\,,
\label{}
\end{equation}
with the lower index $s$ refering to the respective value of the
variable at the sonic surface.
Since, $\tilde R_{\rm s, eq}=v_{\rm s, eq}=1$ we have from the two criticality 
conditions,
\begin{equation}
{GM\over 2 V^{2}_{\rm s, eq}R_{\rm s, eq}}=1
\,,
\label{gm}
\end{equation}
such that
\begin{equation}
\label{vR}
\tilde R_{\rm s} (\psi)  = {1 \over v^2_{\rm s} (\psi) }
\,.
\end{equation}

We are interested in obtaining the flow at large distances
from the central source. Therefore we shall take the distribution of
the velocity and mass flux at infinity as the input parameters of the
problem and introduce the parameter 
$$\xi(\psi)={V_{\infty}(\psi)\over V_{\rm \infty , eq}}
\,.
$$
Taking into account the above two criticality conditions we have,

\begin{equation}
{v^{2}_{\rm s} (\psi) \over 2}+{v_{\rm s}^{2}(\psi)
\over \gamma -1}-2v_{\rm s}^{2}(\psi)={
v_{\rm \infty , eq}^{2}\xi^{2}(\psi)\over 2}
\,,
\end{equation}
which gives
\begin{equation}
v_{\rm s}^{2}(\psi)=\xi^{2}(\psi)
\,,
\end{equation}
and
\begin{equation}
v_{\rm \infty , eq}^{2}={5-3\gamma\over \gamma-1}
\,.
\end{equation}
Note that $v_{\rm  \infty , eq} \geq 1$ only if $\gamma\leq 3/2$. 
The enthalpy function $\tilde Q (\psi )$ can be calculated in
terms of $\xi (\psi)$ and $\dot m (\psi )$ from
Eq. (\ref{mass}) evaluated at the sonic surface and
Eqs. (\ref{vR}) - (\ref{crit1}),
\begin{equation}
\gamma \tilde Q(\psi ) = {\xi^2 (\psi) \over \tilde {\rho_{\rm s}}^{\gamma - 1}(\psi )}
= {\xi^{5-3\gamma}(\psi)\over \dot m^{\gamma-1}(\psi)}
\,.
\end{equation}
From Eqs. (\ref{mass}) - (\ref{vR}), the density at the critical surface
is given in terms of $\xi (\psi )$  and $\dot m (\psi )$,
\begin{equation}
\label{rhoxim}
\tilde \rho_{\rm s} (\psi) = {\xi^3 (\psi) \dot m (\psi )}
\,.
\end{equation}

The Bernoulli equation in dimensionless variables has the form
\begin{equation}
{v^{2}\over 2}+{ \xi^{5- 3 \gamma} (\psi) \over \gamma-1}
\left({ v \tilde R^2}\right)^{(1- \gamma )}
-{2\over \tilde R}={5-3\gamma\over (\gamma-1)}{\xi^{2}(\psi)\over 2} 
\,.
\label{ber3}
\end{equation}

The above Bernoulli equation determines the plasma flow $v (\tilde R;
\psi )$ along the prescribed radial magnetic field, once
the polytropic index $\gamma$ and the distribution of the asymptotic
velocity $\xi(\psi)$ are given. Then, Eq. (\ref{mass}) gives the density
$\tilde \rho = \dot m /v \tilde R^2$ once the mass flux
$\dot m(\psi)$ accross the poloidal streamlines is given.
Note that in order to finally calculate the physical variables $V$ and
$\rho$ we need in addition, as input parameter of the problem, the
equatorial sound speed $V_{\rm s, eq}$ while $\rho_{\rm s, eq}$ can be 
calculated from the given $\dot m_{\rm eq}$.

Finally, consider the initial radial magnetic field
\begin{equation}
B=B_{\rm s, eq}\left({R_{\rm s, eq}\over R}\right)^{2} = 
{B_{\rm s, eq} \over \tilde R^{2}}
\,,
\end{equation}
where $B_{\rm s, eq}$ is the magnetic field at the equatorial sonic transition.
To define a dimensionless magnetic field, $\tilde B$, we need to
normalize $B$ to some characteristic value $B_{\rm c}$ which we choose to be
given by the condition 
%\begin{equation}
$B_{\rm c}^{2}= 4\pi \rho_{\rm s, eq}V_{\rm s, eq}^{2}$. 
%\,.
%\end{equation}
The dimensionless magnetic field $\tilde B \equiv {B/ B_{\rm c}}$ then has 
the form
\begin{equation}
\tilde B  =
{B_{\rm s, eq}\over\sqrt{4\pi\rho_{\rm s, eq}}V_{\rm s, eq}}
\left({R_{\rm s, eq}\over R}\right)^{2}=
{V_{\rm A, s, eq}\over V_{\rm s, eq}}{1 \over \tilde R^{2}}
\,,
\end{equation}
where $V_{\rm A, s, eq}={B_{\rm s, eq}/ \sqrt{4\pi\rho_{\rm s, eq}}}$
is the Alfv\'enic velocity at the equatorial sonic point. Evidently,
the strength of the initial dimensionless magnetic field is controlled
by the magnitude of the ratio of the Alfv\'en and sound speeds at the
equatorial sonic distance, ${V_{\rm A, s, eq}/ V_{\rm s, eq}}$.

\section{The time-dependent stellar wind problem}

To obtain a stationary solution of the problem in the nearest zone of the
star containing the sonic critical surface, it is needed to solve the
complete system of the time-dependent MHD equations and look for an
asymptotic stationary state.
Then, the plasma flow in a gravitational field with the thermal pressure
included is described by the following set of the familiar MHD equations,

\begin{equation}
{\bf B_{\rm p}}={\nabla\psi\times {\bf \hat {\varphi}}\over r}
\,,
\label{psi}
\end{equation}

\begin{equation}
{\partial\psi\over\partial t} =-V_{r}{\partial\psi\over\partial
r}-V_{z}{\partial\psi\over\partial z}
\,,
\end{equation}

\begin{equation}
{\partial \rho \over\partial t}=- {1\over r} {\partial\over \partial
r} (\rho rV_{r}) - {\partial \over \partial z}( \rho V_{z})
\,,
\end{equation}

\begin{equation}
{\partial B_{\varphi}\over\partial t}={\partial
\over\partial z} (V_{\varphi}B_{z}-V_{z}B_{\varphi})
-{\partial \over\partial r} (V_{r}B_{\varphi}-V_{\varphi}B_{r})
\,,
\end{equation}

\begin{eqnarray}
{\partial V_{\varphi}\over\partial
t}&=&-{V_{r}\over r} {\partial \over \partial r}(rV_{\varphi})
-V_{z}{\partial V_{\varphi}\over\partial z}
\nonumber \\
& & +{1\over 4\pi\rho}\left( B_{r}{\partial \over r\partial r}(rB_{\varphi})
+B_{z} {\partial B_{\varphi}\over\partial z}\right) 
\,,
\end{eqnarray}

\begin{eqnarray}
{\partial V_{z}\over\partial t}&=&-V_{r}{\partial V_{z}\over\partial
r}-V_{z}{\partial V_{z}\over\partial z}
-{1\over\rho}{\partial P\over\partial z}-{GMz\over R^{3}}
-{1\over 8\pi\rho r^{2}}\times \nonumber \\
& & {\partial
\over\partial z} (rB_{\varphi})^{2} 
 -
{B_{r}\over 4\pi\rho}\left({\partial B_{r}\over\partial z}-{\partial
B_{z}\over\partial r}\right) 
\,,
\end{eqnarray}

\begin{eqnarray}
%\displaystyle{
{\partial V_{r}\over\partial t}&=&-V_{r}{\partial V_{r}\over\partial
r}-V_{z}{\partial V_{r}\over\partial z}
-{1\over\rho}{\partial P\over\partial r}-{GMr\over R^{3}}- {1\over 8\pi\rho r^{2}} 
\times 
\nonumber \\
& &
{\partial
\over\partial r} (rB_{\varphi})^{2} 
+ {V_{\varphi}^{2}\over r}+
{B_{z}\over 4\pi\rho}\left({\partial B_{r}\over\partial z}-{\partial
 B_{z}\over\partial r}\right)  
\,,
\end{eqnarray}

\noindent
where we have used cylindrical coordinates $(z, r, \varphi )$
and the magnetic field B has a poloidal magnetic flux
denoted by $\psi (z, r)$.
In the simulation we assumed a polytropic equation of state, so that
the relationship of pressure and density in the plasma is 
$P=Q(\psi)\rho^{\gamma}$, at any point along the flow.

A correct solution of the problem requires a specification of the
appropriate boundary conditions at the base surface. 
%Previous studies have used the stellar surface,
%placed below the slow mode critical surface, in order to specify the
%appropriate boundary conditions (Washimi and Shibata 1993, Sakurai 1985).
%However, in 
In this paper our main intention is to compare our results with observations
at large distances from the source. In other words, we are interested in 
the asymptotic properties of the
plasma flow in stellar winds. %Unfortunately, our understanding of the
%various processes contributing to the acceleration of winds close to
%their base is inadequate to reproduce correctly the flow in the zone
%between the stellar base and the slow mode critical surface.  For
%example, in the case of the solar wind it has been realized that the
%acceleration of the outflow has basically a nonthermal origin,
%i.e., Alfv\'en waves, etc. With this limitation in mind, it 
It is reasonable
to start our numerical integration at a boundary surface $R_o$
placed just downstream of the slow mode critical surface.
In this way we avoid the problems connected with the correct description
of the acceleration of the flow below the slow mode critical surface.
Nevertheless, our boundary sourface will be placed below the Alfv\'en and
fast mode critical surfaces.
The correct passage of the physical solution from these two critical
surfaces will yield the appropriate values of the toroidal component of
the magnetic field which is important in controlling the process of
outflow collimation.

Then, the appropriate physical conditions at the boundary surface of
integration are in dimensional variables:\\
1. The density of the plasma at $R=R_o$  kept constant in time, 
although it may depend on the colatitude $\theta$, $\rho = \rho_o(\theta )$.\\
2. The total plasma speed $V_o(\theta )$ in the corotating frame of
reference at $R=R_o$, kept constant in time, 
although it may also depend on the colatitude $\theta$,
$V_{(r,o)}^2 + V_{(z,o)}^2 + (V_{(\varphi , o)} - \Omega r_{o})^2 = V_o^2
(\theta )$ where $V_o (\theta )$ is the plasma speed on the surface $R_o$ 
of the initial flow. The value of the initial velocity $V_o (\theta )$ 
was taken such as to yield the observable values of the terminal velocity 
of the wind from a nonrotating star, i.e., typical solar wind velocities 
at 1 AU.\\
3. A constant and uniform in latitude distribution of the magnetic flux
function at $R=R_o$, $\psi = \psi_o$.\\
4. Finally, the continuity of the tangential component of the electric field
across the stellar surface in the corotating frame gives the last condition,
$(V_{(\varphi , o)} - \Omega r_o) B_{(p, o)} - V_{(p, o)}B_{(\varphi , o)}
= 0$.

Recently it was realized by Ustyugova et al. (1999) that the boundary 
conditions at the outer box of simulation are important for obtaining 
the correct physical stationary solution. 
The importance of a correct specification of the 
outer boundary conditions in MHD outflows has been emphasized previously 
in Bogovalov (1996, 1997) where we controlled the position of the outer 
boundary in the region where no signal can propagate from this boundary into
the box of the simulation. For the solution of the full system of equations
on the outer boundary we used only internal derivatives.

\section{Method of numerical solution of the problem of 
stationary plasma flow to large distances from the star}
%the transfield equation in the coordinates $(\psi, \eta )$ and 

The asymptotic solution of the time-dependent problem in the nearest zone
containing the Alfv\'en and fast mode surfaces was used as the boundary
condition for the far zone wherein we have a supersonic stationary flow.
This boundary condition was then used in order to solve the system of
the MHD equations describing the stationary outflow of superfast
magnetosonic plasma. This system of equations consists of
the set of the MHD integrals and of the transfield equation.

%\subsection{%MHD Integrals and t
%The transfield equation in the coordinates $(\psi, \eta )$ and 
%method of numerical solution of the problem to large distances.}

As is well known, the stationary MHD equations admit four integrals. They
are the 
%\begin{description}
%\item[($\alpha$)] The 
ratio of the poloidal magnetic and mass fluxes,  %$F(\psi )$
%\begin{equation}
$F (\psi ) = {B_{\rm p}/ 4\pi \rho V_{\rm p}}$;  
%\,.
%\end{equation}
%\item[($\beta$)]  
the total angular momentum per unit mass, % $L (\psi )$,
%\begin{equation}
$rV_{\varphi}-FrB_{\varphi}={L(\psi)}$; 
%\,.
%\end{equation}
%\item[($\gamma$)] 
the corotation frequency $\Omega (\psi )$ in the
frozen-in MHD condition
%\begin{equation}
$V_{\varphi}B_{p} - V_{\rm p}B_{\varphi} = r B_{\rm p} \Omega (\psi )$ 
%\,.
%\label{freez}
%\end{equation}
%\noindent
%\item[($\delta$)] 
and finally the total energy $E(\psi )$ in the equation for
total energy conservation,
%\begin{equation}
%{V^{2}\over 2}+{\gamma \over \gamma -1} {P\over \rho} -{GM\over R}
%-F(\psi)\Omega (\psi) rB_{\varphi}=E(\psi)
%\,.
%\end{equation}
%\end{description}

The method of the solution of the transfield equation in the super fast
magnetosonic region is described in detail in our previous Paper I. 
An orthogonal curvilinear coordinate system
($\psi, \eta$) is used, formed by the tangent to the poloidal magnetic
field line $\hat \eta = \hat p$ and the first normal towards the center
of curvature of the poloidal lines,
$\hat \psi = \nabla \psi /|\nabla \psi|$.
A geometrical interval in these coordinates can be expressed as
\begin{equation}
(d{\bf r})^{2}=g^2_{\psi}d\psi^{2}+g^2_{\eta}d\eta^{2}+r^{2}d\varphi^{2},
\end{equation}
where $g_{\psi}, g_{\eta}$ are the corresponding line elements, i.e., 
the components of the metric tensor.

According to Landau \& Lifshitz (1975) the equation $T_{\psi; k}^{k}=\rho
{\partial\over\partial\psi} ({GM\over R})$,
where $T^{ik}$ is the energy-momentum tensor, will have the following
form in these coordinates in the nonrelativistic limit,
\begin{equation}
\begin{array}{c}
\displaystyle{
{\partial\over\partial\psi}\left[P+{B^{2} \over
8\pi}\right]- {1\over r}{\partial r\over
\partial\psi}\left[\rho V_{\varphi}^{2} -{B_{\varphi}^{2} 
\over 4\pi} \right]-}\\
\
\displaystyle{
{1\over g_{\eta}} {\partial g_{\eta} \over \partial\psi}
\left[ \rho V_{\rm p}^{2} -{B_{\rm p}^{2} \over 4\pi} \right] =
\rho{\partial\over\partial\psi} ({GM\over R})
}
\,.
\end{array}
\label{transfield}
\end{equation}

%\section{Method of numerical solution of the problem at large distances.}

It is convenient to solve the transfield equation in the system of
the curvilinear coordinates ($\psi, \eta$) introduced above. The unknown 
variables
are $z(\eta, \psi)$ and $r(\eta, \psi)$. Therefore we need to know the
quantities  $r_{\psi}$, $z_{\psi}$, $r_{\eta}$,  $z_{\eta}$, where 
$r_{\eta}=\partial r/\partial\eta$, $z_{\eta}=
\partial z/\partial\eta$, $r_{\psi}=\partial r/\partial\psi$,
$z_{\psi}=\partial z/\partial\psi$ and $g_{\eta}$, $g_{\psi}$.

First, the metric coefficient $g_{\eta}$ is obtained from the above 
transfield equation (\ref{transfield}),
\begin{equation}
 g_{\eta} =\exp{(\int\limits_a^\psi G(\eta,\psi)d\psi)},
 \label{ga}
 \end{equation}
 where
        \begin{equation}
\displaystyle{
         G(\eta,\psi)={{\partial\over\partial\psi}\left[P+{B^{2} \over
        8\pi}\right] - \rho{\partial\over\partial\psi}({GM\over R})
        -{1\over r} {\partial r\over
        \partial\psi} \left[\rho V_{\varphi}^{2}-{B_{\varphi}^{2} \over 4\pi}
        \right] \over \left[ \rho V_{\rm p}^{2} -{ B_{\rm p}^{2} \over
        4\pi}\right]}
}
\,.
          \label{G}
        \end{equation}

The lower limit of the integration in (\ref{ga}) is chosen to be 0
such that the coordinate $\eta$ is uniquely defined.
In this way $\eta$ coincides with the coordinate $z$ where the
surface of constant $\eta$ crosses the axis of rotation.

The metric coefficient $g_{\psi}$ is given in terms of the magnitude of
the poloidal magnetic field,
\begin{equation}
g_{\psi} ={1\over rB_{\rm p} }
\,.
\end{equation}

To obtain the expressions of $r_{\psi}$, $z_{\psi}$,  
$r_{\eta}$,  $z_{\eta}$ 
we may use the orthogonality condition
\begin{equation}
 r_{\eta}r_{\psi}+z_{\eta}z_{\psi}=0
\,,
\label{orto}
\end{equation}
and also the fact that they are related to the metric coefficients
$g_{\eta}$ and $g_{\psi}$ as follows,
\begin{equation}
g^2_{\eta}=r_{\eta}^{2}+z_{\eta}^{2}
\,,~~~~g^2_{\psi}=r_{\psi}^{2}+z_{\psi}^{2}
\,.
\label{galpha}
\end{equation}

Finally, by combining the condition of orthogonality (\ref{orto}) and Eqs. 
 (\ref{galpha}) the remaining values of
$r_{\eta}$, $z_{\eta}$ are obtained,

\begin{equation}
r_{\eta}=-{z_{\psi} g_{\eta} \over g_{\psi}}
\,,~~~~z_{\eta}={r_{\psi} g_{\eta} \over g_{\psi}}
\,,
\label{za}
\end{equation}
with $g_{\eta}$ above defined by the expression (\ref{ga}).
For the numerical solution of the system of equations (\ref{za})
the two step Lax-Wendroff method on the lattice with a dimension equal
to 1000 is used.

Eqs. (\ref{za}) should be supplemented by
appropriate boundary conditions on some initial surface of constant $\eta$.
The equations for $r_{\psi}$, $z_{\psi}$ defining this initial surface
in cylindrical coordinates are,

\begin{equation}
{\partial r\over \partial\psi}={B_{z}\over rB_{\rm p}^{2}},
~~~{\partial z\over \partial\psi}=-{B_{r}\over rB_{\rm p}^{2}}
\,.
\label{rpsi}
\end{equation}

We need to specify  on this surface the integrals $E(\psi), L(\psi),
\Omega(\psi), F(\psi)$ as the boundary conditions for the initial
value problem.
To specify the initial surface of constant $\eta$ and the above integrals,
we use the results of the solution of
the problem in the nearest zone when a stationary solution is
obtained for the time-dependent problem.

\section{Parametrization of the stationary solution}

It is convenient to consider the solution in dimensionless variables.
In the present paper we express the velocities in units of the initial sound 
speed at the sound point on the equator $V_{\rm s, eq}$, all the geometrical 
variables in units of the initial radius of the sound point
on the equator $R_{\rm s, eq}$ and the magnetic field in units of  
$B_{\rm c}=\sqrt{4\pi\rho_{\rm s, eq}} V_{\rm s, eq}$.

In this notation the solution depends on a few parameters.
Among them is the ratio  of the radius of the star to the radius of 
the initial sound point $\tilde R_* = R_*/R_{\rm s, eq}$, the parameters  
\begin{equation}
\beta= \displaystyle{\Omega R_{\rm s, eq} \over V_{\rm s, eq}}
\,,
\end{equation}
and 
\begin{equation}
V_a %= B_{\rm s, eq}/B_c 
={V_{\rm A, s, eq}\over V_{\rm s, eq}}
\,,
\end{equation}
together with the two dimensionless functions $\xi(\psi)$ and $\dot m(\psi)$ which
are equal to 1 in the case of uniform ejection of plasma at the base.

Since we are not interested here in the dependence of the solution on the 
radius of the star, basically the flow is defined by the two parameters 
$\beta$ and $V_a$.

Using the above definition of $\beta$ and the relationships at the critical 
surface, Eqs. (\ref{gm}) - (\ref{vR}), 
this parameter can be expressed through observable parameters as
\begin{equation}
\beta= {\Omega G M\over 2 V_{\infty}^3} ({5-3\gamma\over\gamma-1})^{3/2},
\end{equation} 
where $V_\infty$ is the terminal velocity of the plasma for the nonrotating
star with mass $M$. %Correspondingly,  
%$$
%C_{eq,s}=V_{\infty}({\gamma-1\over 5-3\gamma})^{1/2}, ~~~~
% R_{eq,s}= {G M\over 2 V_{\infty}^2} ({5-3\gamma\over\gamma-1})\,.
%$$ 
The parameter 
\begin{equation}
V_a = {\psi_t\over\sqrt{\dot M R^2_{\rm s, eq} V_{\rm s, eq}}}\,,
\end{equation}
where $\psi_t = \lim_{r\rightarrow\infty} B\cdot r^2$ is estimated on the 
equator of the nonrotating star and $\dot M$ is the total mass loss of the 
nonrotating star.

In Paper I the degree of collimation of the outflow depended critically 
on a parameter $\alpha$ which was defined for cold plasmas as the ratio 
of the corotation speed at the Alfv\'en transition to the initial velocity
of the plasma. However for hot winds where the velocity depends on the 
radial distance we should introduce a more general definition of this parameter 
$\alpha$. 

In general, winds from astrophysical objects  can be driven by a combination 
of several mechanisms of acceleration such as the gradients of thermal pressure, 
Alfv\'en wave and radiation pressures, magnetocentrifugal forces, etc. 
It is natural to characterize the efficiency of the 
magnetocentrifugal forces by the ratio $V_m/V_{\infty,0}$ where 
%\begin{equation}
$V_m = ({\Omega^2 \psi_t^2/\dot M})^{1/3}$% 
%\,,
%\end{equation}
is Michel's (1969) terminal velocity of a plasma accelerated only by
magnetocentrifugal forces in the split-monopole model provided that the initial 
velocity is zero while $V_{\infty,~ 0}$ is the terminal wind speed 
due to all other mechnisms of acceleration. 
In other words, $V_{\infty, ~0}$ is  the terminal velocity of the plasma 
for the nonrotating star. In this case the generalized
$\alpha$ can be defined as
\begin{equation}
\alpha = \left({V_m\over V_{\infty,0}}\right)^{3/2}
\,.
\end{equation} 

In the special case of cold plasma outflow this definition of the parameter $\alpha$ 
coincides with the parameter introduced in our previous Paper I. 
It can be easily shown that for polytropic winds this parameter can be 
expressed as
\begin{equation}
\alpha=\beta V_a ({\gamma-1\over 5-3\gamma})^{3/4}
\,,
\end{equation}
or,
\begin{equation}
\alpha = {\psi_t\Omega\over\sqrt{\dot M} V_{\infty,0}^{3/2}}.
\end{equation}

Magnetic rotators with $\alpha > 1$  shall be called fast magnetic 
rotators and those with $\alpha < 1$ slow magnetic rotators. 
It is useful to note that for a slow magnetic rotator such as the Sun, 
$R_A \approx \psi_t/(\dot M V_{\infty,0})^{1/2}$ 
(MacGregor 1996), and hence $\alpha \approx \Omega R_A/V_{\infty,~0}$.

%figure 3
\setbox1=\vbox{\hsize=7.0 truecm \vsize=7.0truecm
\psfig{figure=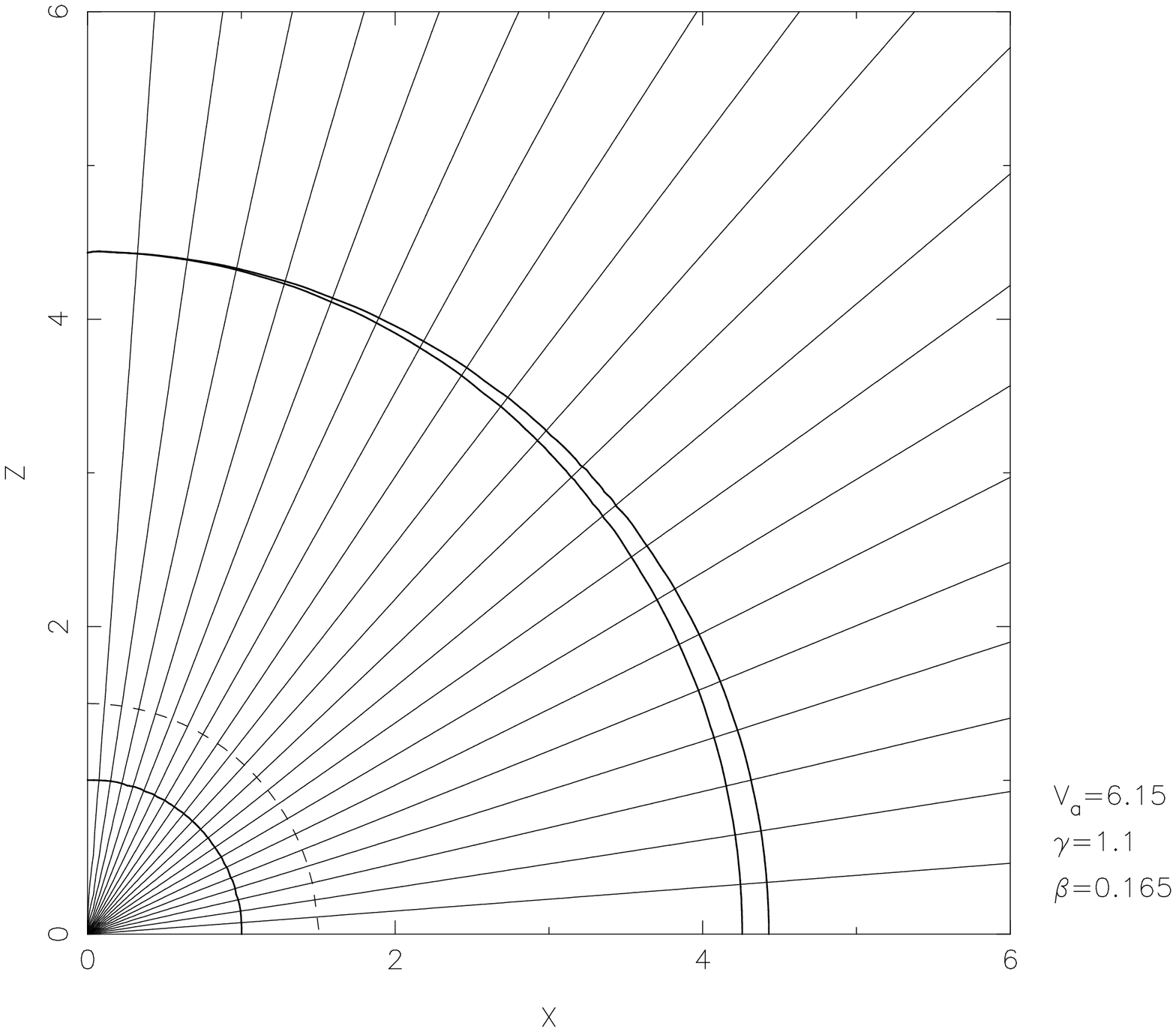,height=7.0truecm,angle=270}}
\setbox2=\vbox{\hsize=7.0 truecm \vsize=7.0truecm
\psfig{figure=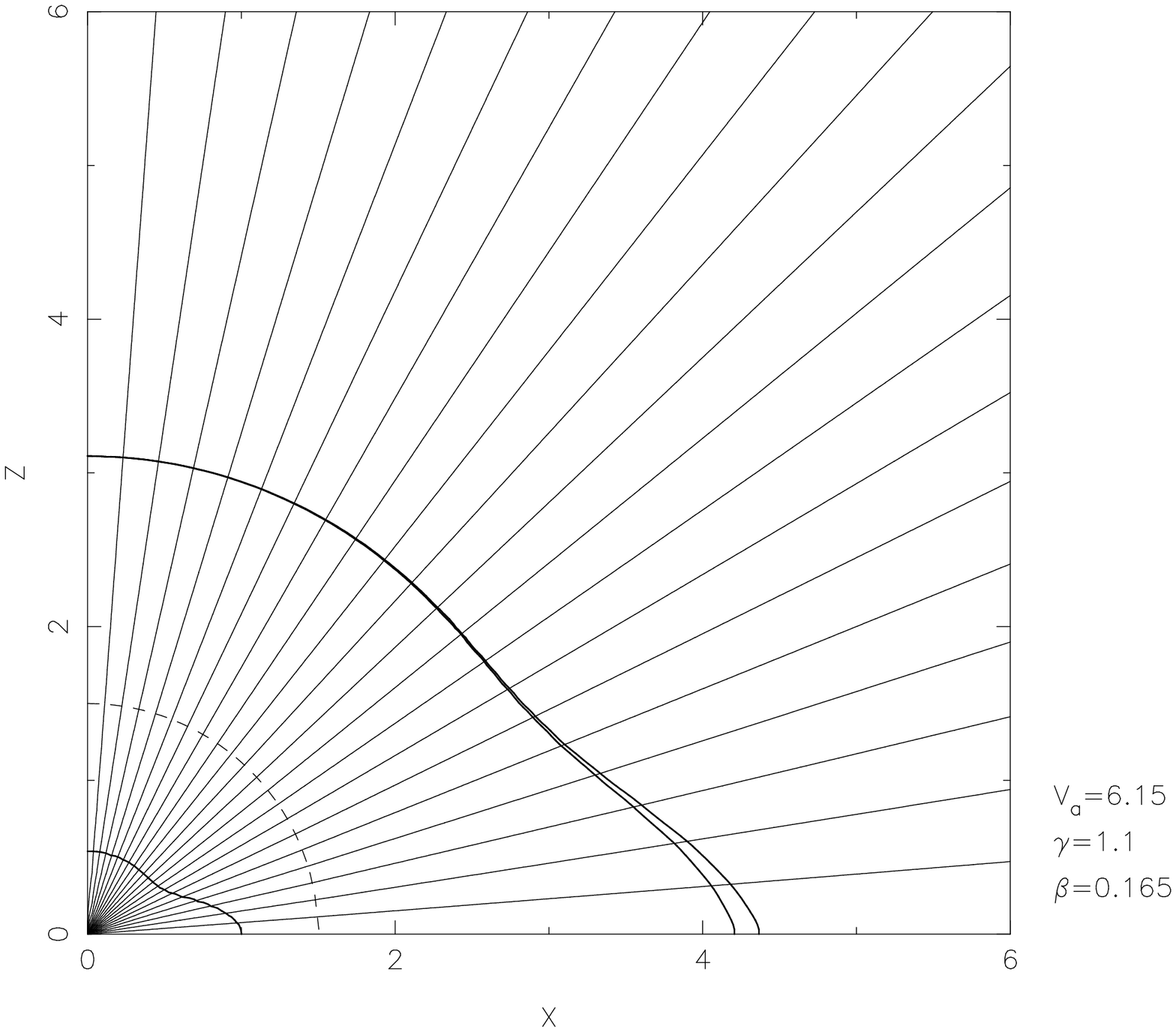,height=7.0truecm,angle=270}}
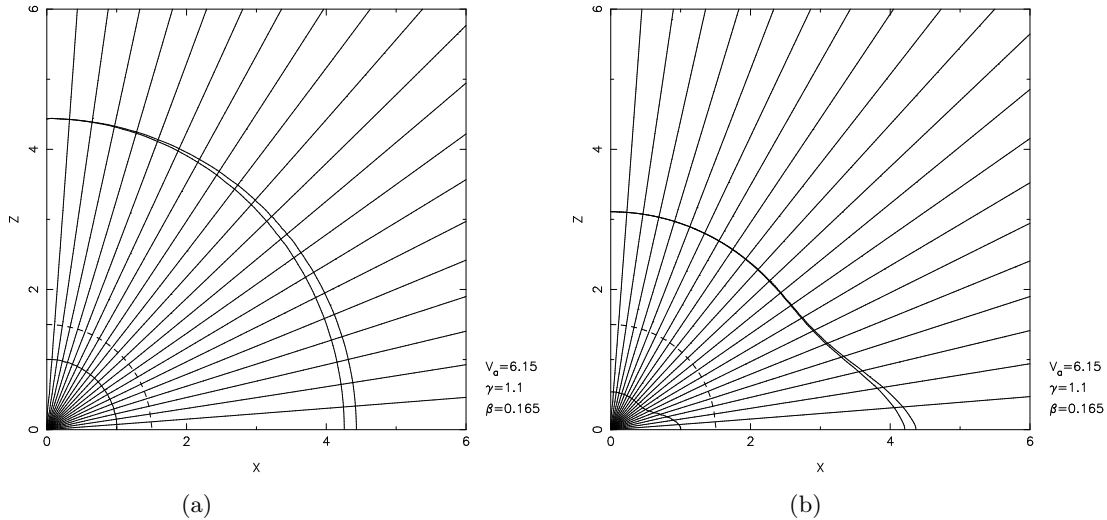
\begin{figure*}%[h]
\centerline{\box1\hspace{0.5cm}\box2\hspace{3.5cm}}
%\vspace{1cm}
\begin{picture}(400,1)
\put(120.0,0.0){(a)}
\put(350.0,0.0){(b)}
\end{picture}
\caption{Shape of poloidal magnetic field lines and streamlines
in the {near zone} of an isotropic (panel a) and anisotropic 
(panel b) solar wind with $\gamma =1.1$, 
$V_a = 6.15$ and $\beta=0.165$.
Spherical distance is given in units of radius of slow
point at $R_{\rm slow} = 8.4 ~R_{\odot}$, with the base radius at
$R=1.5$ (dashed line). Thick lines indicate slow, Alfv\'en
and fast critical surfaces.}
\label{near_sun}
\end{figure*}

%Figure 4
\setbox1=\vbox{\hsize=7 truecm \vsize=7truecm
\psfig{figure=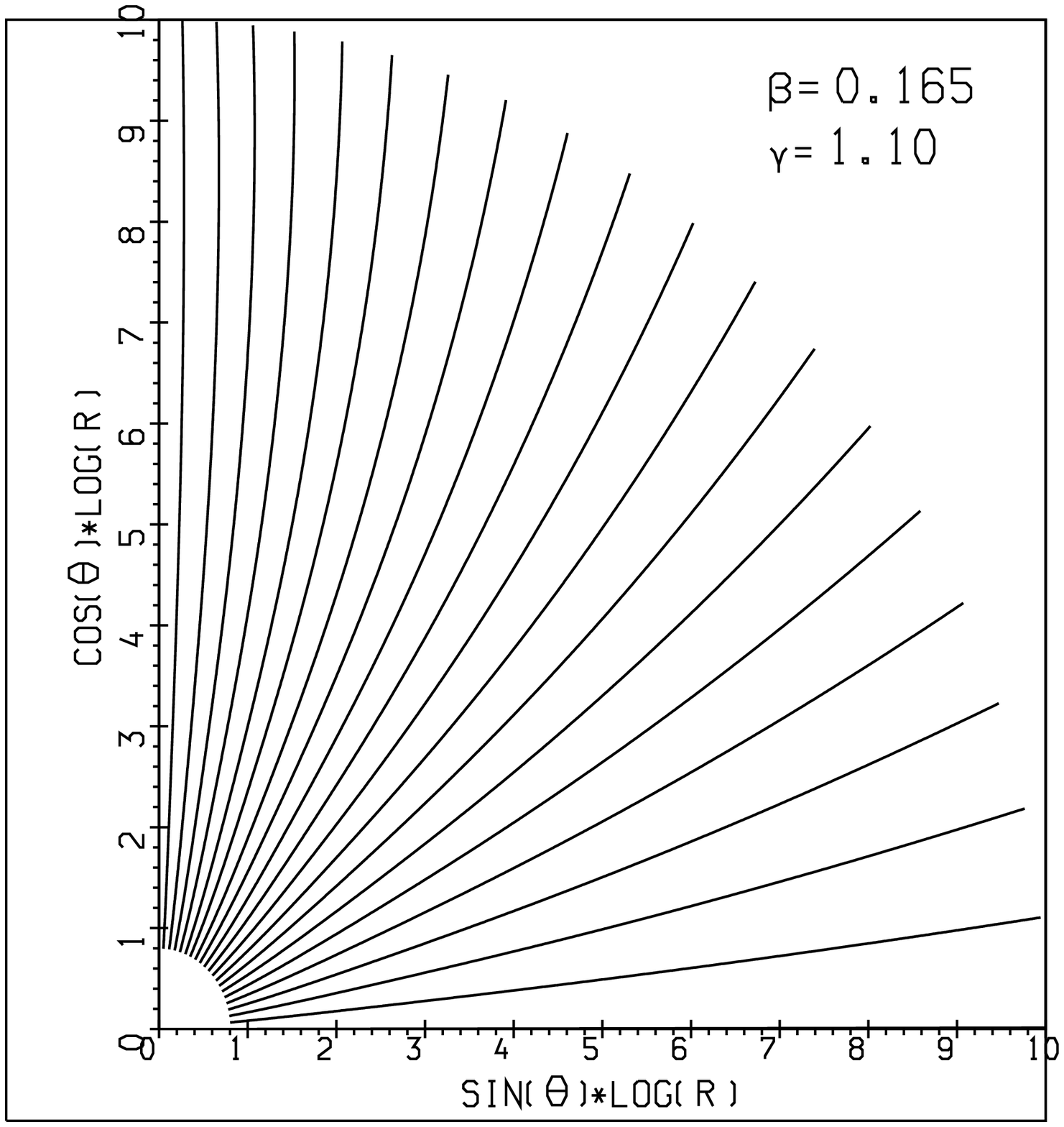,height=7.0truecm,angle=360}}
\setbox2=\vbox{\hsize=7 truecm \vsize=7truecm
\psfig{figure=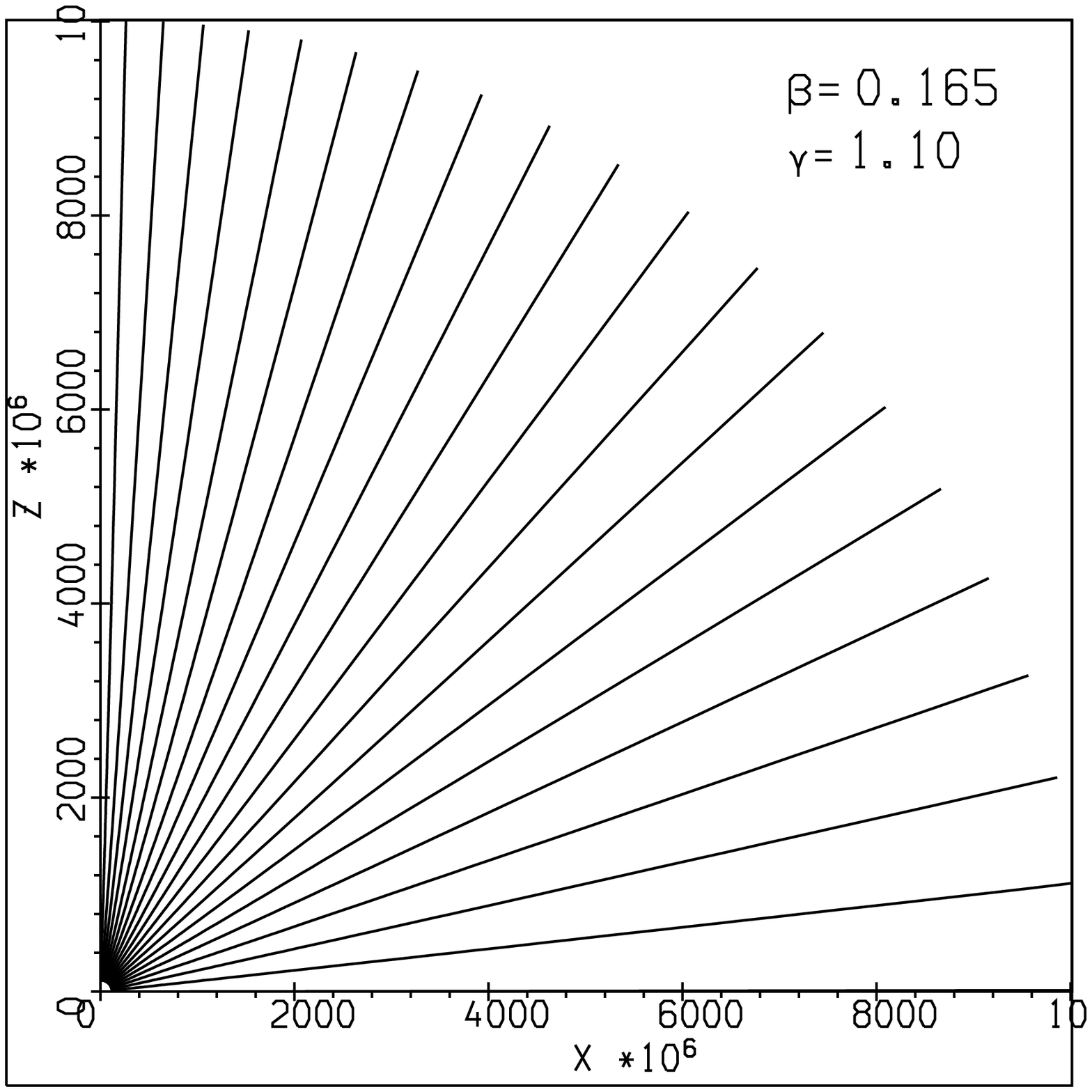,height=7.0truecm,angle=360}}
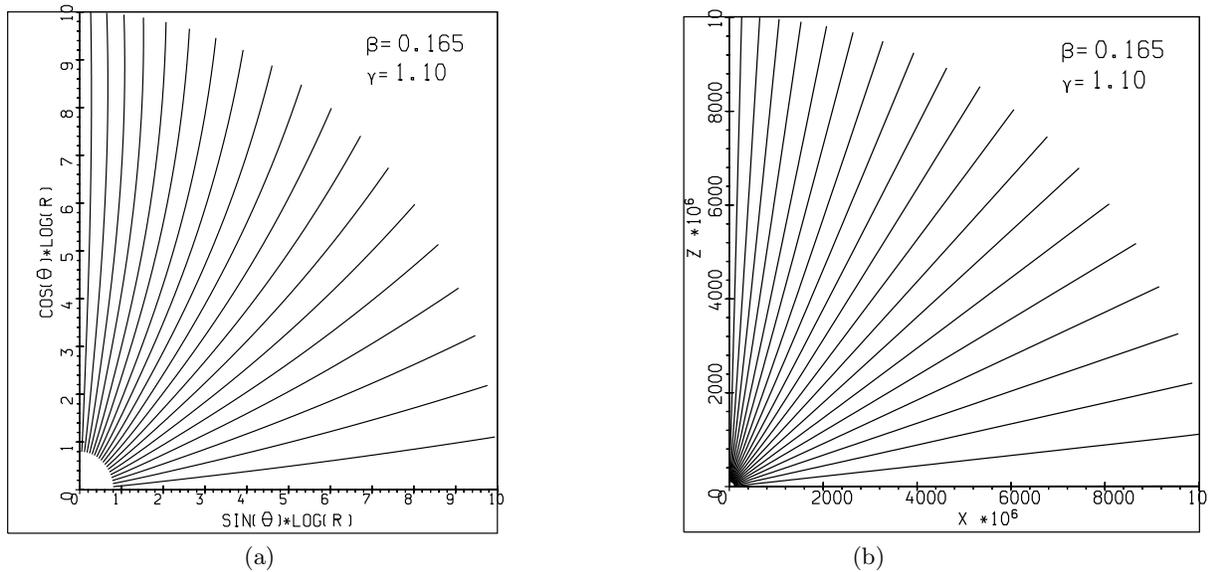
\begin{figure*}%[h]
\centerline{\box1\hspace{2.0cm}\box2\hspace{1.0cm}}
%\vspace{1cm}
\begin{picture}(400,1)
\put(120.0,0.0){(a)}
\put(350.0,0.0){(b)}
\end{picture}
\caption{Shape of poloidal magnetic field lines in the far zone of an 
{isotropic} SW with  $\gamma = 1.1$ and $\beta=0.165$.
In (a) the poloidal field lines are plotted
in a logarithmic scale, which magnifies their slight bending 
towards the axis, while in the linear scale of
(b) it may be seen that collimation is negligible.
}
\label{shape_sun_1.1}
\end{figure*}

A quantitative classification of magnetic rotators on slow and fast 
(Belcher \& MacGregor 
1976) has also been introduced in Ferreira (1997) by using the parameter 
${\Omega R_A/V_A}$, where $V_A$ is the Alfv\'en velocity in the Alfv\'en 
transition located at the radius $R_A$. This parameter
is also less than 1 for slow rotators, but for fast rotators it is of the 
order of 1, since for fast magnetic rotators $V_A \sim \Omega R_A$ 
(Michel 1969). Nevertheless, physically this classification of 
magnetic rotators to fast and slow coincides in both cases. Note also in 
passing that in terms of an energetic criterion for collimation deduced in 
Sauty \& Tsinganos (1994) and Sauty et al. (1999) magnetic rotators are 
analogously classified as efficient (with cylindrical asymptotics) and 
inefficient (with radial asymptotics). 
%\newpage

\section{Results for the isotropic and nonisotropic solar wind}

The most often used magnetized 
polytropic solar wind model is the classical one 
proposed by Weber \& Davis (1967) where the poloidal magnetic field 
and stream lines are radial (see also MacGregor 1996). This model 
reproduces the observed properties of the low speed streams at 1 AU 
within the observed fluctuations (Charbonneau 1995). 

We will choose in our analysis the polytropic index to the value 
$\gamma = 1.1$ such that the wind is heated. 
The spherical distance will be expressed in units of the distance
of the slow magnetosonic point $R_{\rm slow} \approx 
8.4 ~R_{\odot}$ and the velocity in units of the slow magnetosonic 
speed there, $V_{\rm slow} \approx 106$ km/s. The {\it fast} magnetosonic 
transition occurs at $R_{\rm fast} %\approx 29.5 ~ R_{o} 4.5 ~ R_{\rm slow} 
\approx 38 ~ R_{\odot}$ 
where the fast magnetosonic speed is $V_{\rm fast}  \approx 230$ km/sec.
Note that for slow magnetic rotators like our Sun, the slow magnetosonic 
speed almost coincides with the sound speed and the Alfv\'en critical point 
with the fast magnetosonic critical point. In particular, at the axis 
the Alfv\'en and fast transitions coincide but at the equator the 
Alfv\'en transition occurs earlier as in Figs. \ref{near_sun} 
(see also Paper I and Keppens \& Goedbloed 1999) 

%Most of the acceleration of the SW takes place within the distance 
%of the slow magnetosonic point. Deviations from radiality, as e.g., 
%flaring of the magnetic field lines in coronal holes, also take place at 
%$R\leq R_{slow}$. In this study  we shall exclude in our 
%modelling the acceleration region $R\leq R_{slow}$ and put the base of the 
%simulation at a distance slightly above the slow magnetosonic point, 
%$R= 1.5 ~ R_{slow}$. 
%In fact, by placing the base above the sonic transition, we model 
%the almost radial configuration of the solar magnetic field as shown 
%in Fig. \ref{schematic}. 
%More complex magnetic configurations like coronal streamers 
%will not be taken into account in this first approximation of the problem 
%where we assume that the magnetic field is unipolar in each hemisphere. 

In Fig. \ref{near_sun}a we plot the shape of the poloidal magnetic field 
lines and streamlines in the near zone of a wind which is isotropic at the 
base with $\gamma =1.1$, $V_a = 6.15$ and $\beta=0.165$. With these parameters 
$\alpha =0.12$, i.e., in our terminology the Sun is a slow rotator. 
Careful inspection 
of this figure shows that the flow is very slightly collimated to the 
axis of rotation. The solution in the far zone is presented in 
Fig. \ref{shape_sun_1.1}. 

%%##

In Fig. \ref{shape_sun_1.1}a the poloidal field lines of the SW are plotted 
in a logarithmic scale, which magnifies their slight bending towards the 
axis. This logarithmic scale extends to the huge distance of $10^{10} 
R_{\rm slow}$, i.e, about $4\times 10^{8}$ AU $\approx 60$ light years. 
This figure shows that the solar wind is indeed collimated toward  
the axis of rotation.
But this collimation is indeed very weak. Fig. \ref{shape_sun_1.1}b
shows the same field in a linear scale which shows that a jet is still 
not formed even at these huge distances.  
%Figure 5
\setbox18=\vbox{\hsize=6 truecm \vsize=6truecm
\psfig{figure=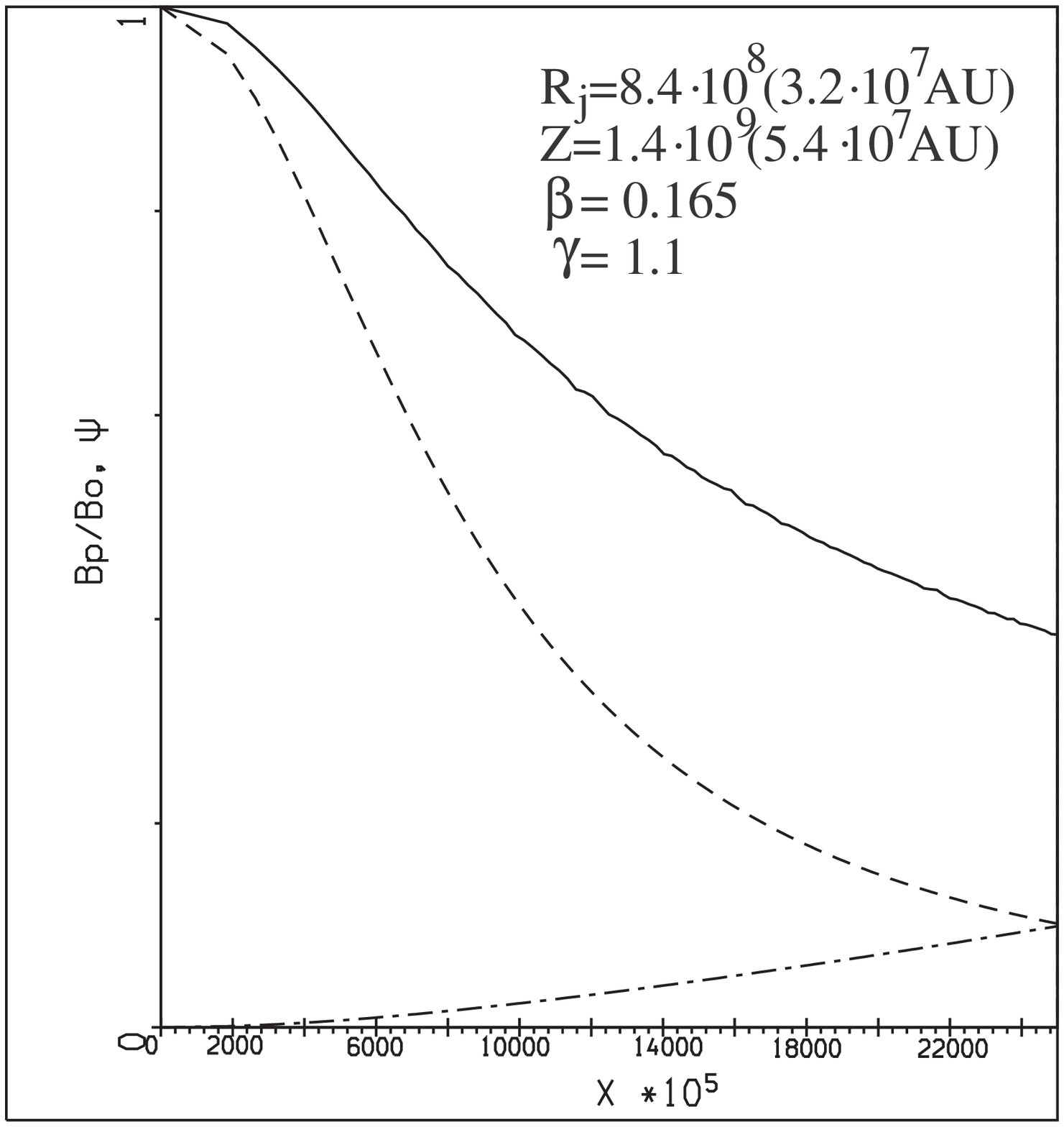,height=6.0truecm,angle=0}}
\setbox19=\vbox{\hsize=6 truecm \vsize=6truecm
\psfig{figure=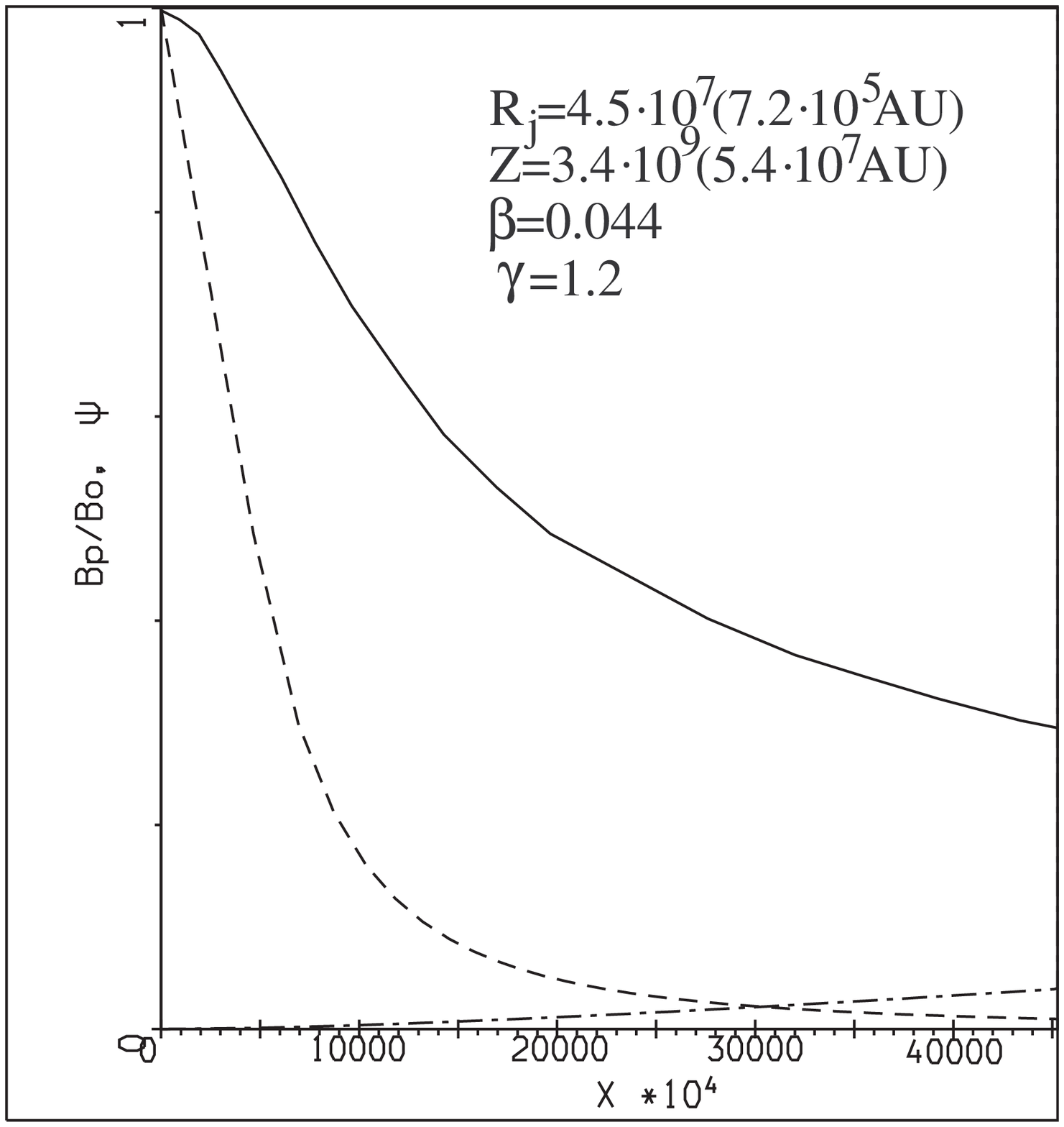,height=6.0truecm,angle=360}}
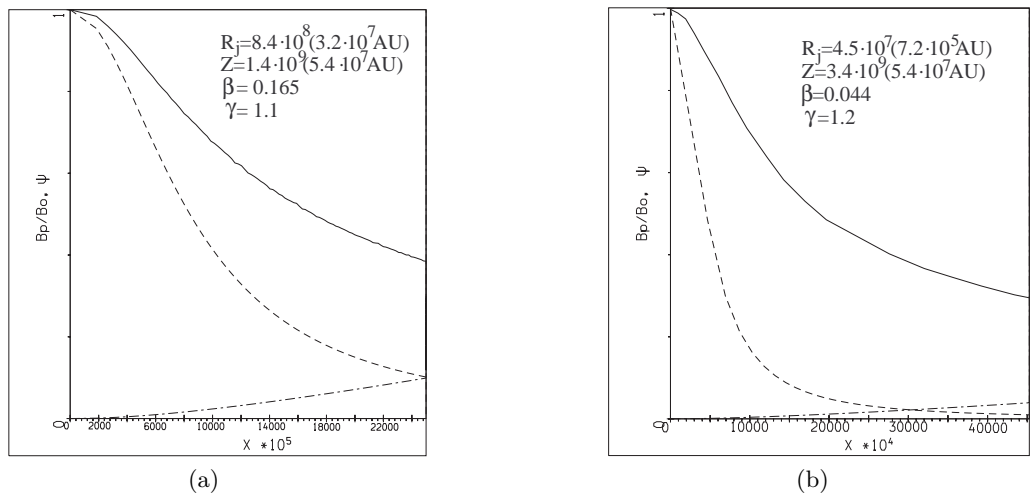
\begin{figure*}%[h]
\centerline{\box18\hspace{2.0cm}\box19\hspace{1.5cm}}
%\vspace{1cm}
\begin{picture}(400,1)
\put(120.0,0.0){(a)}
\put(350.0,0.0){(b)}
\end{picture}
\caption{Variation with dimensionless cylindrical distance 
$X=r/R_{\rm slow}$ of enclosed magnetic flux $\psi(X)$ (dot-dashed) and magnetic
field $B_p (X)/B_o$ (solid) %and density $n(r)/n_o$ for an isotropic wind 
for $\beta=0.165$ and $\gamma=1.1$ (panel a) and 
$\beta=0.044$ and $\gamma=1.2$ (panel b). Dashed line indicates the 
analytically predicted values of $B_p (X)/B_o$.}
\label{jetsun}
\end{figure*}

As is shown in Bogovalov (1995), the dependence of the magnitude of the 
poloidal magnetic field and density on the cylindrical distance 
$r$ becomes rather simple if we assume for convenience that 
the MHD integrals $E(\psi), L(\psi), F(\psi), \Omega(\psi)$ and the 
terminal velocity of the jet $V_{j}$ are constants and do not depend 
on $\psi$ and also that $V_{j} \gg V_{A}(0)$, where
$V_{A}(0)$ is the Alfv\'enic velocity at the axis of rotation $r = 0$. 
In such a case, an approximate estimate of the dependence of the magnetic 
field on $r$ is given in terms of the magnetic field and the 
density of the plasma on the axis of the jet, $B_{\rm p}(0)$ and $\rho(0)$,  
respectively,
\begin{equation}
{B_{\rm p}(r)\over B_{\rm p}(0)}={\rho (r)\over \rho(0)}=
{1\over 1+(r /R_{j})^{2}} \,.
\label{jr}
\end{equation}
The radius of the core of the jet $R_{j}$ is given in terms of 
the sound and Alfv\'en speeds along the jet's axis $V_{\rm s}(0)$ and 
$V_{A}(0)$, 
\begin{equation}
R_{j}=\sqrt{\left(1+{V_{\rm s}(0)^{2}\over V_{A}(0)^{2}}\right)}{V_{j}\over\Omega}
\,,
\end{equation}

Hence, the poloidal magnetic field and density remain practically constant 
up to axial distances of order $R_j$ and then they decay fast like $1/r^2$ 
outside the jet's core.

The comparison of this theoretical prediction with the characteristics of the
solar wind is shown in Fig. \ref{jetsun}. 
There is a remarkable discreapancy between them which tells us that even 
at these unrealistically huge distance (the SW wind is presumably 
terminated earlier) the jet has still not 
formed. The reason is that the collimation of the wind occurs  
logarithmically with distance (Paper I) 
provided that the jet is not formed in the nearest zone, as it happens 
for fast rotators. For the solar parameters, there isn't enough
radial distance to form the jet.

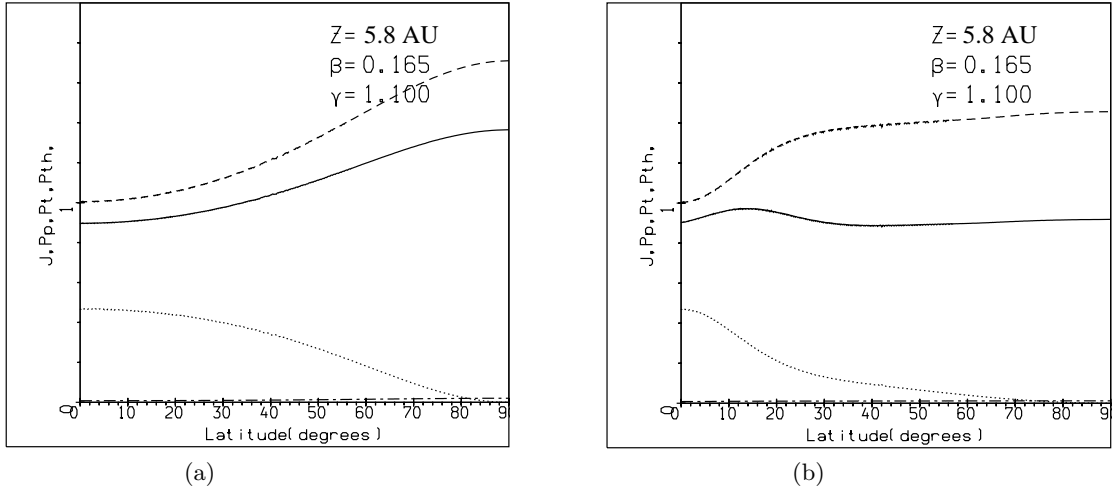
\begin{figure*}%[h]
\centerline{\box33\hspace{2.0cm}\box34\hspace{1.5cm}}
%\vspace{1cm}
\begin{picture}(400,1)
\put(120.0,0.0){(a)}
\put(350.0,0.0){(b)}
\end{picture}
\caption{Distribution with latitude of characteristics of 
isotropic (panel a) and anisotropic (panel b) solar wind at 5.8 AU 
for $\beta=0.165$ and 
$\gamma = 1.1$. Dashed lines indicate thermal pressure 
$P_{\rm th}(\theta )/P_{\rm th}(\theta =0)$, 
solid lines mass flux $J (\theta ) =\rho v R^2$, 
dotted lines pressure of toroidal magnetic field 
$P_{\rm t}(\theta )/P_{\rm th}(\theta =0)$ and  
dotted-dashed lines pressure of poloidal magnetic field 
$P_{\rm p}(\theta )/P_{\rm th}(\theta =0)$.}
%\begin{figure}
%\centerline{\hspace{-2.5cm}\box33\hspace{1.5cm}\box34}
%\vspace{1cm}
%\centerline{\box35}
\label{iso_aniso}
\end{figure*}

In spite of the absence of a jet core in the solar wind, the lines of 
the plasma flow are certainly bent to the axis of rotation. 
And some observable effects can arise due to this bending. 
If the base density is  isotropic, the SW mass efflux increases 
with the latitude $\theta$ 
because of the magnetic focusing by about 20\% from the equator to 
the pole (Fig. \ref{iso_aniso}a) at a distance from the Sun of about 5.8 AU.
Approximately at this distance the interplanetary $L_{\alpha}$ emission 
is formed, as observed by the SWAN instrument on board of SOHO. 
This theoretical anisotropy is in contradiction with the measurements 
of the anisotropy of the SW at large distance from the Sun by  
SWAN. This discreapency however, can be eliminated if we take into 
account some initial anisotropy in the SW.

To study in more detail the effect of the focusing of the SW, we 
peformed  calculations
for a more realistic model of the SW including some initial anisotropy 
of the wind at its base.  
%Figure 6
\setbox33=\vbox{\hsize=6 truecm \vsize=6truecm
\psfig{figure=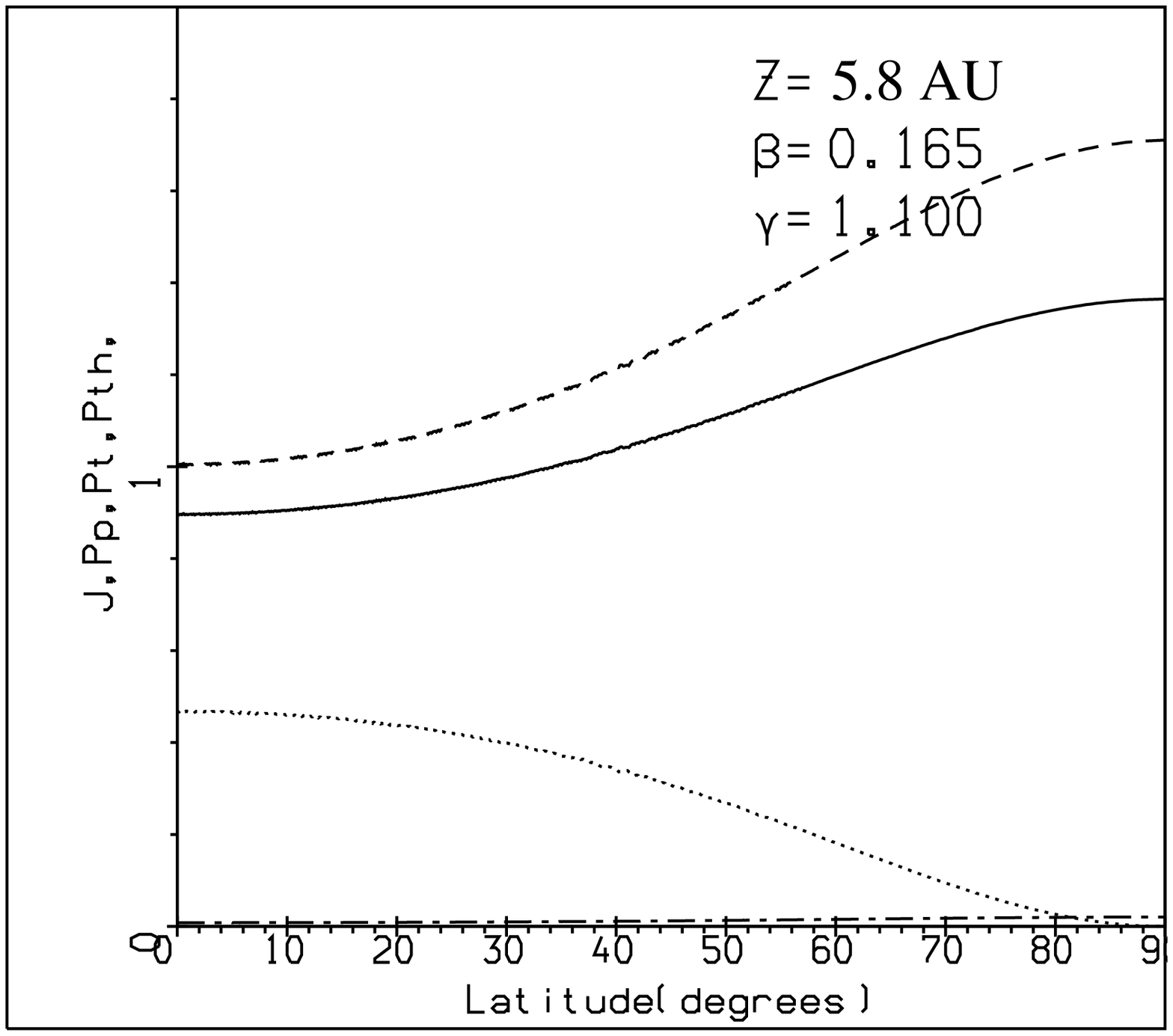,height=6.0truecm,angle=270}}
\setbox34=\vbox{\hsize=6 truecm \vsize=6truecm
\psfig{figure=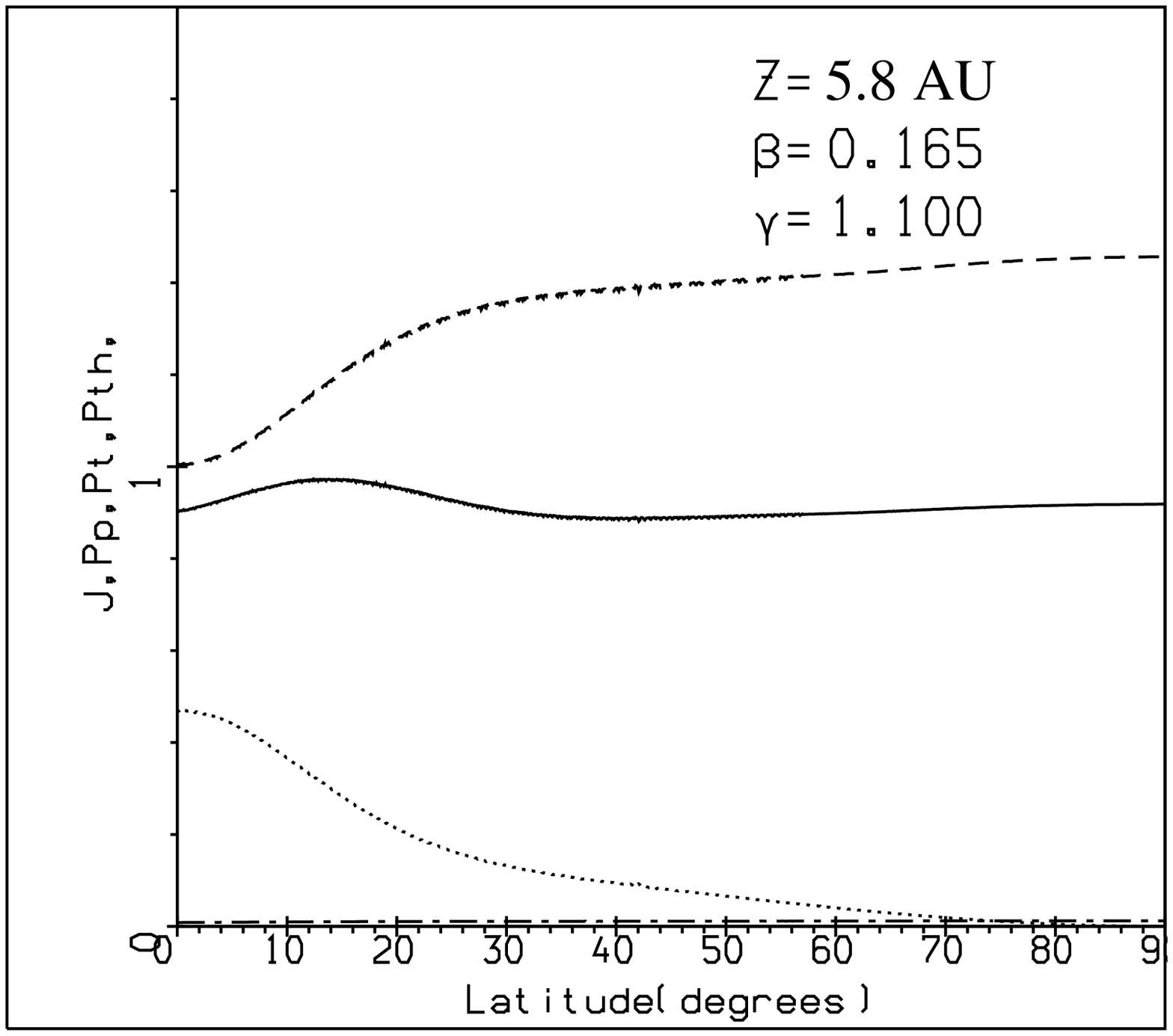,height=6.0truecm,angle=270}}

In Fig. \ref{near_sun}b the shape of the streamlines 
and Alfv\'en/fast critical surfaces is shown 
in the near zone of a wind which 
is latitudinally anisotropic in density and speed at the reference 
base at $R=1.5 ~R_{\rm s, eq}$. 
For this simulation we used the latitudinal dependence of an exact MHD 
solution for the solar wind (Lima et al. 1997, Gallagher et al. 1999) with 
\begin{equation}\label{1}
\rho(\theta)=
{\rho_0}\left({1+\delta \sin^{2\epsilon}\theta}\right)\,, 
\; 
V_r(\theta)= V_{o}
\sqrt{{1+\mu \sin^{2\epsilon}\theta\over 1+\delta \sin^{2\epsilon}\theta}}
\,,
\end{equation}
\noindent
where $V_0$, and $\rho_0$ correspond to the reference values
of the flow speed and density and the distribution depends on the three 
parameters $\delta$, $\mu$ and $\epsilon$. 
The values of $\epsilon$, $\delta$ and 
$\mu$ have been chosen in Lima et al. (1997) as to best reproduce the 
observed values of the wind speed at various latitudes, as obtained recently 
by Ulysses (Goldstein et al. 1996). Their deduced values which we adopted in 
this study are $\delta = 1.17$, $\mu = -0.38$ and $\epsilon = 8.6$. 
A similar density enhancement about the ecliptic and an associated 
increase of the wind speed around the poles is also found in recent  
MHD simulations as well (Keppens \& Goedbloed 1999). 

The distribution of the mass efflux at the distance of 5.8 AU for the
anisotropic at the base SW is shown in Fig. \ref{iso_aniso}b.
The effect of the magnetic focusing almost totally disappers at 
high latitudes.
Near the equator the excess  of the mass efflux remains remarkable in 
the region below 30 degrees, although 
evidently the mass efflux decreases below 15 degrees.
Therefore these results are in reasonable agreement with the 
distribution of the mass efflux of the SW which is deduced by SWAN 
at distances 5-7 AU.

The drastical decrease of the effect of the focusing of the solar wind in the
anisotropic case can be naturally explained by the larger velocity of the
SW at high latitudes. According to observations by ULYSSES, 
the velocity at high latitudes is almost twice higher than the velocity 
at the equator (Feldman et al. 1996). It follows
from the equation of motion Eq. (\ref{transfield}) that the 
curvature radius $R_c$ of the streamlines in the poloidal plane 
at large distances can be estimated as (Paper I)
\begin{equation}
{1\over r^2}
{\partial\over\partial\psi}{r^2 B_{\varphi}^2 \over 8\pi}  \sim 
{\rho V_p^2\over rR_c B_p} 
\,.
\end{equation}
The toroidal magnetic field at large distances can be estimated from 
the frozen in 
condition as $B_{\varphi}=-(r\Omega/V_p)B_p$. In this case we have
 \begin{equation}
{1\over R_c} \sim {1\over 8\pi\rho V_p^2 r^3}{\partial\over\partial\theta}
\left ({r\Omega\over V_p}\right )^2 B_p^2 r^2\,.
\end{equation}

The latitudinal distributon of $\rho V_p$ is almost constant with 
an $ \sim 15\%$ increase only near the equator. Therefore the curvature 
radius depends on $V_p$ as
\begin{equation}
{1\over R_c} \sim {1\over V_p^3}
\,,
\end{equation}
provided that the mass flux density is fixed.
Due to this strong dependence of the effect of collimation on 
the velocity of the plasma, the focusing practically disappears at 
high latitudes where the velocity is almost twice the velocity near 
the equator and the distribution of the mass efflux
is in pretty good agreement with the observed one. The decrease of 
the mass efflux below $15^o$ cannot be found by the present SWAN 
data analysis since it was assumed in this analysis that the mass 
efflux can only monotonically increase with decreasing latitude.  

Up to now we have discussed the results of calculations for 
the polytropic wind with $\gamma=1.1$. The flow with this polytropic 
index is almost isothermal and approximates 
the SW near the Sun, up to several dozens of solar radii. 
But at larger distances the effective polytropic index should increase 
and at infinity it should go to the Parker value $3/2$. 
At a first glance, it seems reasonable to expect that
the focusing of the plasma will be stronger for a higher polytropic index. 

To study this possibility we also performed calculations for the 
isotropic solar wind with a polytropic index $\gamma=1.2$, a reasonable 
value at the distance of several AU (Weber 1970).  
The change of the polytropic index results to a change of the other parameters
$\beta$ and $V_a$ so that the total magnetic flux, mass flux and terminal 
velocity of the plasma remain constant as in the case 
with $\gamma=1.1$. In particular, for $\gamma=1.2$ we have 
$\beta=0.044$ and $V_a=11.9$. With these values of $\beta$ and $V_a$ 
the parameter $\alpha$ remains constant.

\begin{figure}[h]
\centerline{\psfig{file=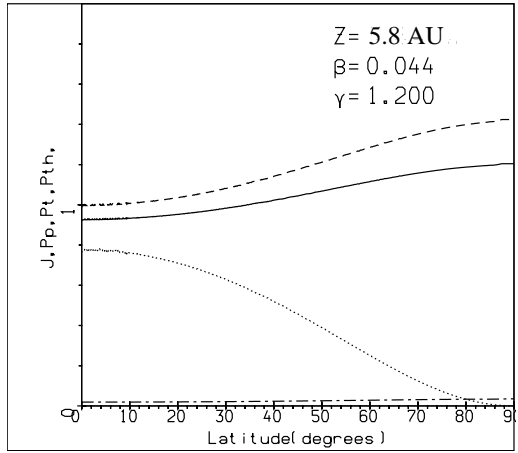,height=6.0truecm,angle=270}}
\caption{Distribution with latitude of the characteristics of the isotropic 
solar wind at 5.8 AU for $\beta=0.044$ and 
$\gamma = 1.2$. Dashed line indicates thermal pressure 
$P_{\rm th}(\theta )/ P_{\rm th}(\theta =0)$, 
solid line mass flux $J (\theta ) =\rho v R^2$, 
dotted line pressure of toroidal magnetic field 
$P_{t} (\theta )/P_{\rm th}(\theta =0)$ and  
dotted-dashed line pressure of poloidal magnetic field 
$P_{\rm p}(\theta )/P_{\rm th}(\theta =0)$.}
\label{gamma1.2}
\end{figure}

The distribution of the mass flux for $\gamma=1.2$ is shown in 
Fig. \ref{gamma1.2}.

%Figure 7

It follows from this figure that the focusing of the flow to the axis 
of rotation becomes even smaller at distances below 5.8 AU than it 
does for $\gamma=1.1$. 
This is explained by the fact that the higher is the polytropic index, 
the less is the variation of the plasma velocity with the distance. 
With a fixed terminal velocity, the wind with $\gamma=1.2$ has a 
higher velocity at the base than the wind with $\gamma=1.1$.
But we already have discussed above the dependence of the focusing 
on the velocity of the flow. It follows from this discussion that  
the focusing is less for the wind with higher velocity. Therefore the 
focusing of the wind at several AU is maximal for $\gamma=1.1$.
Since we have agreement of theory and observations of the solar wind 
anisotropy by SWAN for this polytropic index, it is expected that 
the wind with a higher polytropic index will not provide a higher 
level of collimation which could be inconsistent with the observations.

The dependence of the focusing effect on the polytropic index becomes 
opposite at very large distances where the plasma velocity has already 
achived the terminal value. In this case it becomes 
important how fast the thermal pressure which tends to decollimate 
the plasma falls down with distance.
The higher the polytropic index is,  the faster the pressure falls down 
with the distance. Therefore at very large distances we should expect 
stronger collimation of the plasma for higher values of the polytropic 
index. This tendency is indeed found. Fig.  \ref{jetsun} demostrates 
a comparison of the characteristics of the flow near the axis of 
rotation for polytropic indices 1.1 and 1.2. It is seen that the wind for
$\gamma=1.2$ is stronger collimated  although the formation of the jets 
is also not completed at these huge distances.

%\vfil\newpage
%{\bf Figure 8}
%\setcounter{figure}{4}
\setbox1=\vbox{\hsize=7.0 truecm \vsize=7.0truecm
\psfig{figure=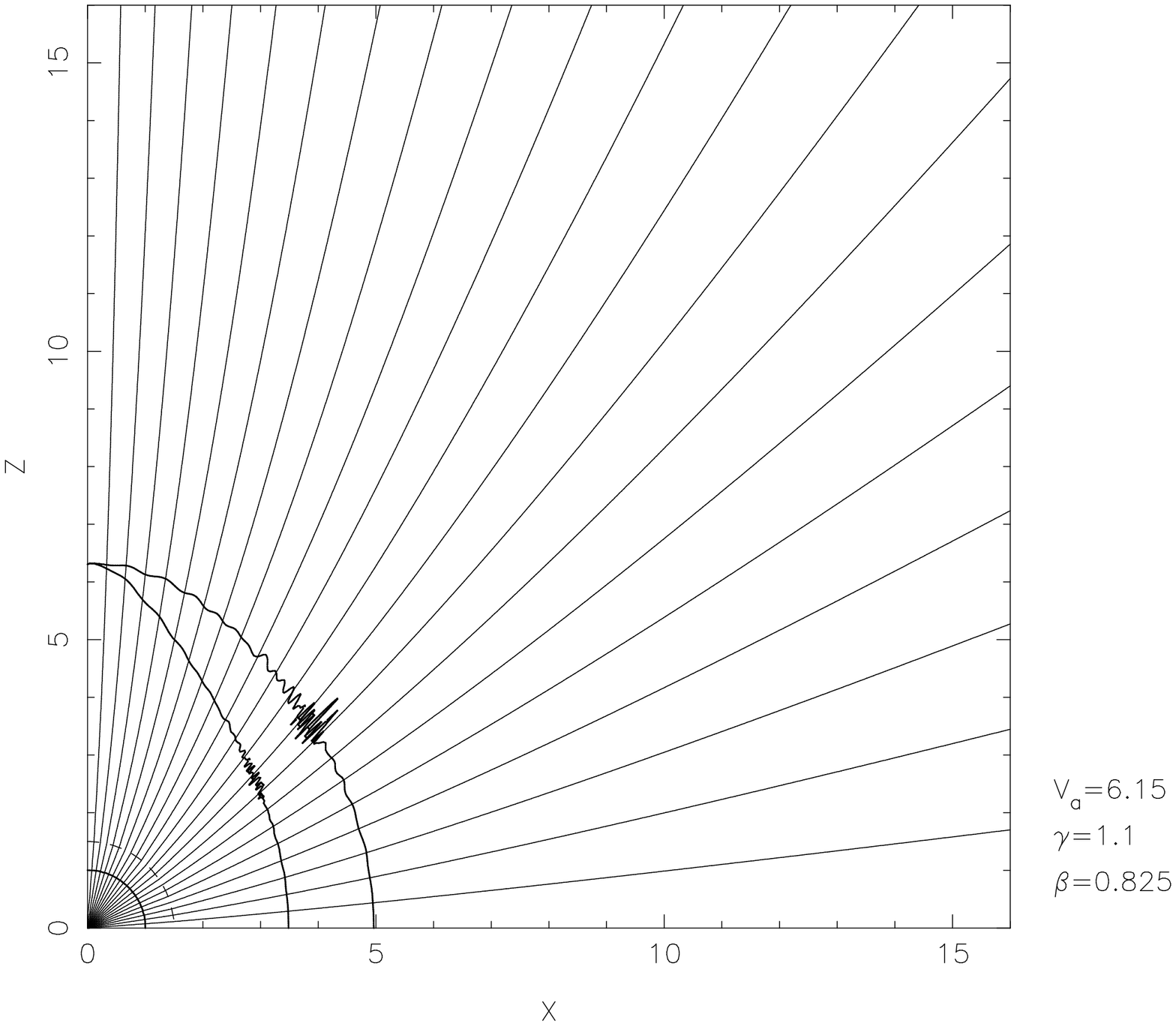,height=7.0truecm,angle=270}}
\setbox2=\vbox{\hsize=7.0 truecm \vsize=7.0truecm
\psfig{figure=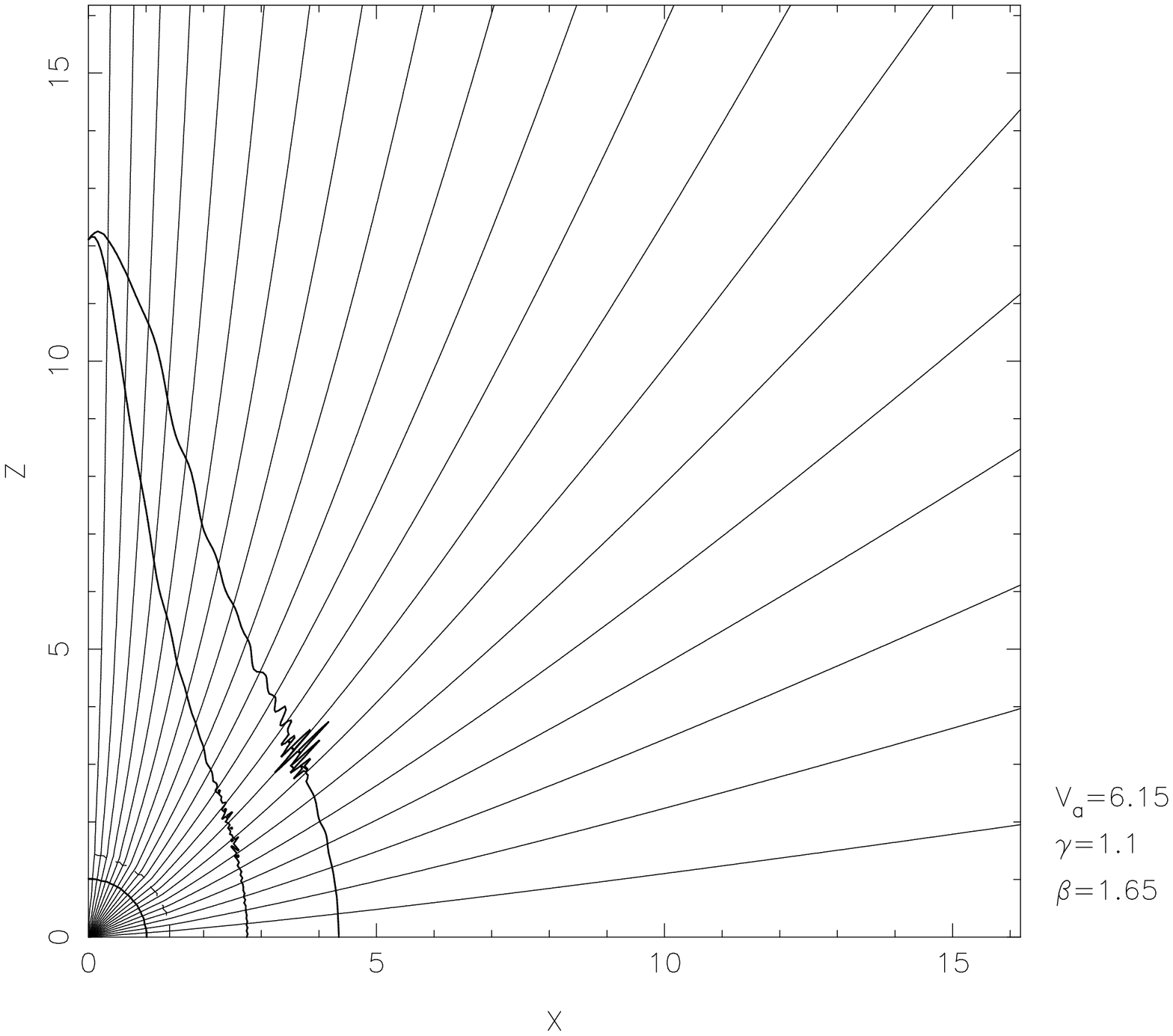,height=7.0truecm,angle=270}}
\begin{figure*}%[h]
\centerline{\hspace{-1.0cm}\box1\hspace{0.5cm}\box2\hspace{3.0cm}}
%\vspace{-0.5cm}
\begin{picture}(400,1)
\put(120.0,0.0){(a)}
\put(330.0,0.0){(b)}
\end{picture}
\caption{Shape of poloidal magnetic field lines/streamlines
in near zone of isotropic wind from a star rotating 5 times
($\beta=0.825$, panel a) and 10 times faster than the Sun, 
($\beta=1.65$, panel b) for $\gamma =1.1$ and 
$V_a = 6.15$.
Spherical distance is in units of radius of slow
point at $R_{\rm slow} = 8.4 ~R_{\odot}$, with the base radius at
$1.5R_{\rm slow}$ (dashed line).
The initial nonrotating monopole magnetic field has a spherical Alfv\'en
surface at $4.5R_{\rm slow}$.
Thick lines indicate slow, Alfv\'en and fast critical surfaces.
}
\label{near5_10}
\end{figure*}
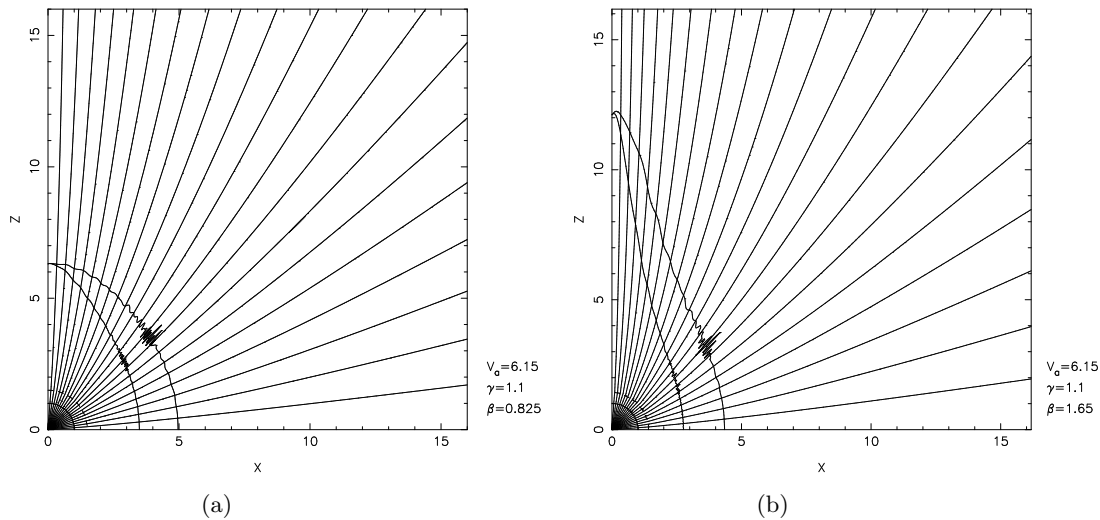

\section{Stellar winds from magnetic rotators faster than the Sun}

As we have seen in the previous section, the Sun is a relatively slow 
magnetic rotator with $\alpha <1$. A plausible senario for the Sun (and 
similar low mass stars) is that it has lost a large fraction of its 
angular momentum via magnetized a outflow while in its 
youth it was in a state of higher rotation, similar to the observed high 
rotation states of young stars (Bouvier et al. 1997) corresponding to larger 
values of $\alpha$. It is interesting then to examine the change of the 
shape of the magnetic field of a stellar magnetic rotator more efficient 
that the Sun, in the context of our simple modelling.  
For convenience and comparison with the present era Sun, we may  
keep constant the parameters of the polytropic index and sound 
speed. In rapid rotators, the magnetic flux also increases roughly in 
proportionality to the rotation rate $\Omega$ (Kawaler 1988, 1990).
For simplicity let us neglect this increase and  
assume that the ratio of the Alfv\'en and sound speed remains the same, 
$V_a = 6.15$, as in the previous example, but the 
parameter $\beta$ increases by 5 and 10 times, from $\beta$ = 0.165 
($\alpha=0.121$) to 
$\beta$ = 0.825  ($\alpha=0.6$) and then to $\beta$ = 1.65 ($\alpha=1.21$).  

We shall again consider in our numerical experiment that the star has 
a radial magnetic field and at t=0 rotation starts which via the Lorentz 
forces distorts this magnetic field. After some time, a final equilibrium 
state is reached (Figs. \ref{near5_10}) where the poloidal magnetic 
field and the plasma density are increased along the axis because of the 
focusing of the field lines towards the pole. 
A test that the steady state solution is reached is the constancy of 
the MHD integrals of motion, $E(\psi), L(\psi), F(\psi)$ and $\Omega(\psi)$.  
Note that the wiggles appearing at the midlatitude in Figs. \ref{near5_10} 
are due to the step-like function representation of the surface of the star 
in the cylindrical coordinate system used in our calculations. This artifact  
is unavoidable in this system of coordinates and is also present in the 
calculations reported in Washimi \& Shibata (1993).   

The shape of the poloidal field lines is shown in the near zone 
in Fig. \ref{near5_10}. Collimation of the plasma to the axis of rotation 
already close to the source is evident. As in the cold plasma case of 
Paper I, the 
elongation of the subsonic region along the axis of rotation is present.
The most surprising result is that this elongation occurs faster with 
an increase of the parameter $\alpha$ than it does for a cold plasma. 
It is easy to compare the flow in the nearest zone for $\alpha=1.2$ 
shown in Fig. \ref{near5_10}b with the corresponding figure for similar 
$\alpha$ presented in Fig. 2 of Paper I. It appears that although the 
shape of the field lines is the same, the subsonic
region for the cold plasma remains  almost spherical at this parameter 
in contract to the shape of the subsonic region for the hot plasma. 
The physics of this interest behaviour is as follows. 
 
The Alfv\'en transition at a given point of the Z-axis occurs when 
$V=(B_p/4\pi\rho V)B_p$. The ratio $B_p/\rho V$ remains constant along 
a field line.
Therefore the right hand side in this expression increases with collimation
as $B_p$. In the cold plasma limit $V$ is constant and the displacement 
of the Alfv\'en point down the flow in the cold plasma occurs only due to 
an increase of $B_p$.
In the hot plasma case, the collimation also modifies the velocity $V$. 

%Figure 9
\begin{figure}[h]
\centerline{\psfig{file=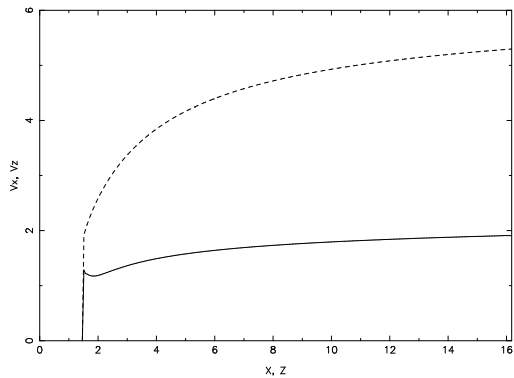,height=5.5truecm,angle=270}}
\caption{Flow speed at the axis (solid) and the equator (dashed line) 
for wind from a star rotating 10 times faster than the Sun,
with  $\gamma = 1.1$ and $\beta=1.65$.
}
\label{velocity}
\end{figure}

This modification is shown in Fig.  \ref{velocity}. 
The solid line 
shows the velocity at the axis and the dashed line shows the velocity 
at the equator. Collimation results in an decrease of the
thermal pressure gradient. Therefore the thermal acceleration of the 
plasma becomes less efficient. The velocity of the plasma even slightly 
decreases due to the gravitation force.
%This means that in the hot plasma case the left hand side of the equation 
%decreases also because of the focusing of the flow, something which did 
%not happen in the cold plasma case. 
Therefore the Alfv\'en surface  in the 
hot case elongates along the axis of rotation with an increase of 
$\alpha$, faster than it does in the cold plasma case. 
It is clear that at some value of $\alpha$ specific for every flow, 
the subsonic region near the axis will be elongated to infinity in 
the Z-direction.
In the hot plasma this happens at smaller $\alpha$ than it does in the 
cold plasma case. This subfast region should be certainly unstable. 
%in relation to helical perturbations (Bateman 1980). 
Therefore, the 
plasma flow at these conditions cannot be stationary. Those jets with  
$\alpha \gg 1$ eventually should become turbulent with properties differing  
from the properties of the stationary jets found for $\alpha < 1$. 
The physics of these jets should be examined in a separate study. 

\begin{figure*}%[h]
\centerline{\box1\hspace{2.0cm}\box2\hspace{1.0cm}}
%\vspace{1cm}
\begin{picture}(400,1)
\put(120.0,0.0){(a)}
\put(350.0,0.0){(b)}
\end{picture}
\caption{Shape of poloidal magnetic field lines for a wind from a star
rotating 10 times faster than the Sun,
with  $\gamma = 1.1$ and $\beta=1.65$.
In (a) the poloidal field lines are plotted
in a logarithmic scale, which artificially magnifies their bending 
towards the axis, while in the linear scale of
(b) a higher degree of collimation in comparison to the 
corresponding case of Fig. 4b  may be seen.
}
\label{shape5_10}
\end{figure*}
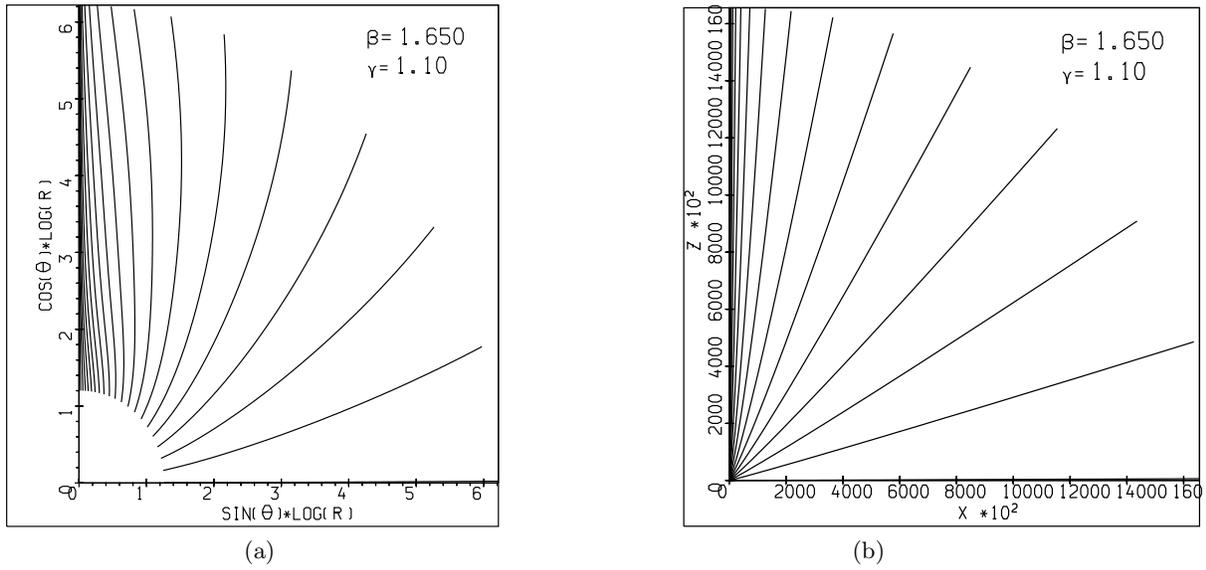

In Fig. \ref{shape5_10} the shape of the poloidal fieldlines is shown 
for the case of the rapid rotator. Their bending towards the rotation 
axis is evident not only in the logarithmic plot of Fig. (\ref{shape5_10}a)   
but also in the linear plot of Fig. \ref{shape5_10}b where a tightly
collimated jet is formed already at a relatively small distance from the
source.
%\newpage
%Figure 10
\setbox1=\vbox{\hsize=7 truecm \vsize=7truecm
\psfig{figure=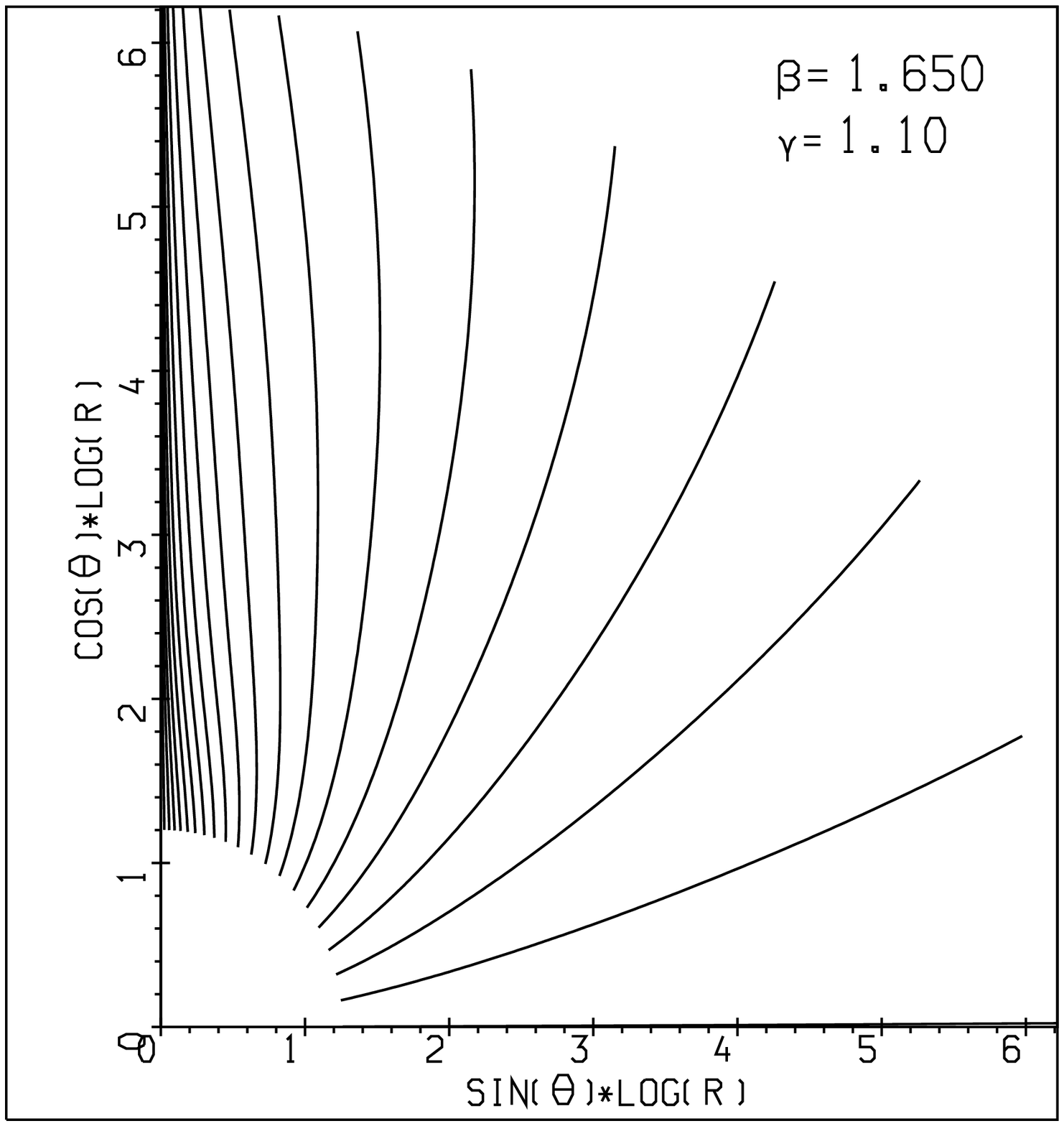,height=7.0truecm,angle=360}}
\setbox2=\vbox{\hsize=7 truecm \vsize=7truecm
\psfig{figure=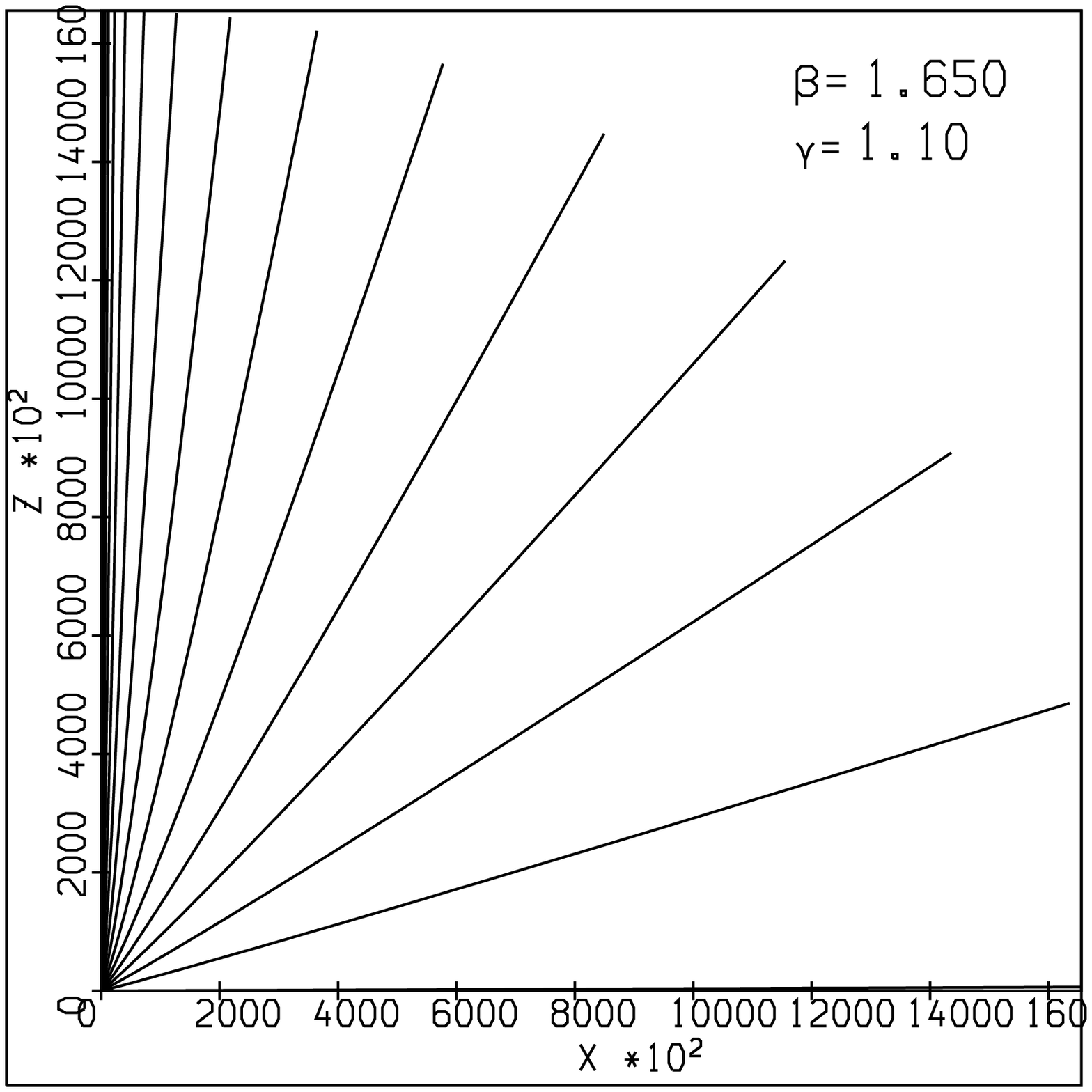,height=7.0truecm,angle=360}}

%Fig. \ref{radius} shows a comparison of the dependence of the density 
%and poloidal magnetic field of the plasma with the distance to 
%the axis of rotation obtained analytically (Bogovalov 1995) and in numerical
%calculations.  There is a rather good agreement between them. 

\begin{figure*}%[h]
\centerline{\box1\hspace{2.0cm}\box2\hspace{1.0cm}}
%\vspace{1cm}
\begin{picture}(400,1)
\put(120.0,0.0){(a)}
\put(350.0,0.0){(b)}
\end{picture}
\caption{Variation with dimensionless cylindrical distance 
$X=r/R_{\rm slow}$ 
of enclosed magnetic flux $\psi(X)$ (dot-dashed) and magnetic
field $B_p (X)/B_o$ (solid) %and density $n(r)/n_o$ for an isotropic wind 
from a moderate stellar magnetic rotator with $\beta=0.825$ (panel a) and a 
faster magnetic rotator with $\beta=1.65$ (panel b). Dashed line indicates the 
analytically predicted values of $B_p (X)/B_o$. }
\label{radius}
\end{figure*}
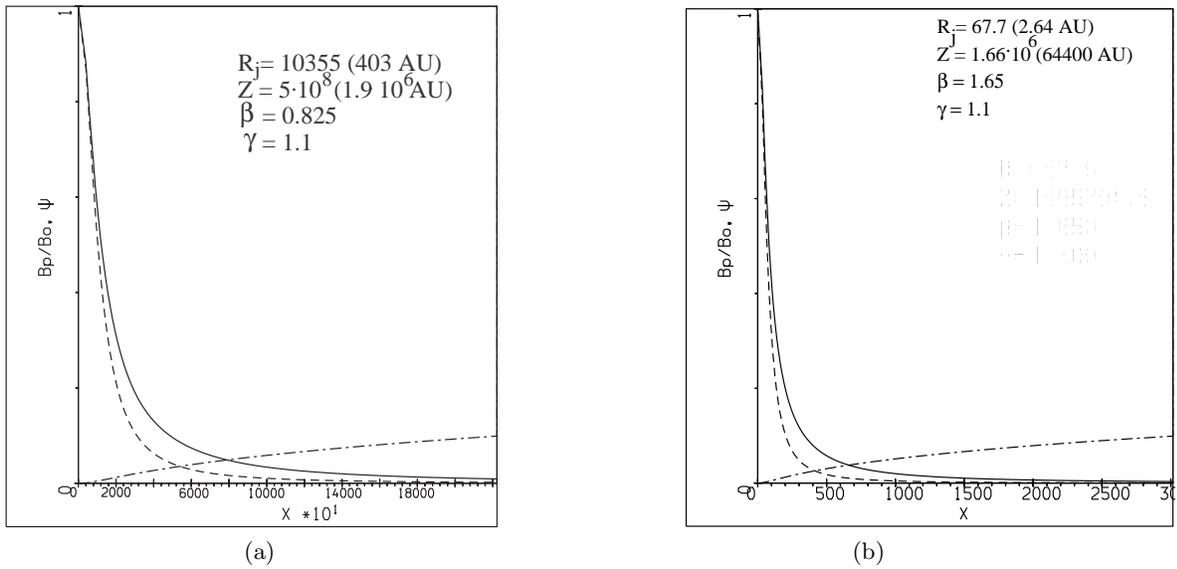

In Figs. \ref{radius} the poloidal component of the magnetic field
is plotted together with the magnetic flux enclosed by a cylindrical 
distance $X$. The poloidal magnetic field $B_p (X)/B_o$
(solid line) is given in units of its reference value $B_o$,
corresponding to the magnetic field at the symmetry axis $r=0$
and some reference height $Z_o=5\cdot 10^8$ for $\beta=0.825$ 
(Fig. \ref{radius}a) and $Z_o=1.66\times 10^6$ for $\beta=1.65$ 
(Fig. \ref{radius}b). The asymptotic regime of the jet 
is achived at these distances. The dashed curve gives the analytically 
predicted solution for the poloidal magnetic field $B_p (X)/B_o$, Eq. (42). 
Note that the agreement between the calculated and analytically predicted 
values of the radius of the jet is  pretty good. It follows from 
this comparison that for these parameters we indeed 
achieve the distance where the jets are really formed.

%{\bf Figure 11}
%\setcounter{figure}{6}
\setbox1=\vbox{\hsize=7.0 truecm \vsize=7.0truecm
\psfig{figure=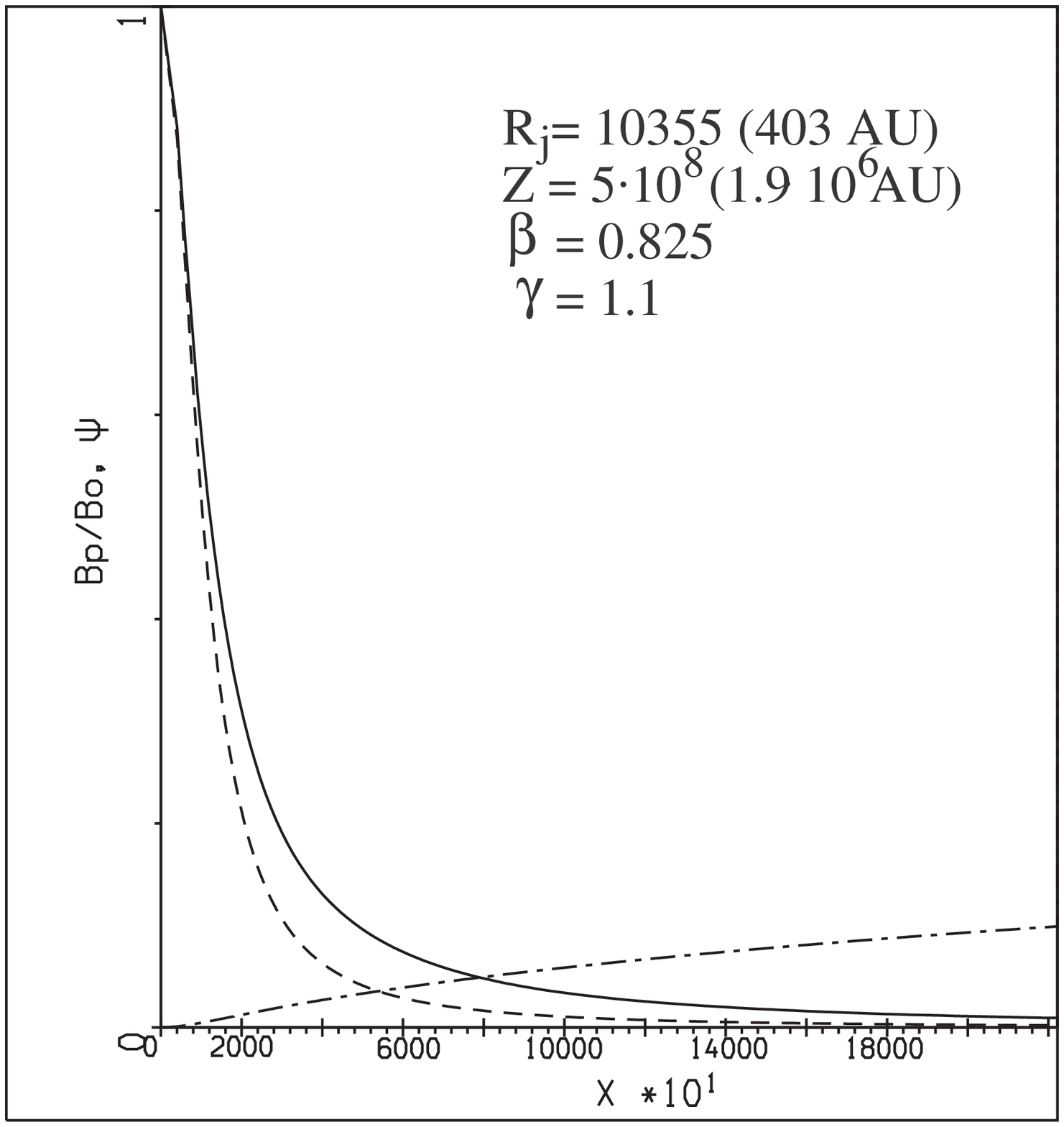,height=7.0truecm,angle=0}}
\setbox2=\vbox{\hsize=7.0 truecm \vsize=7.0truecm
\psfig{figure=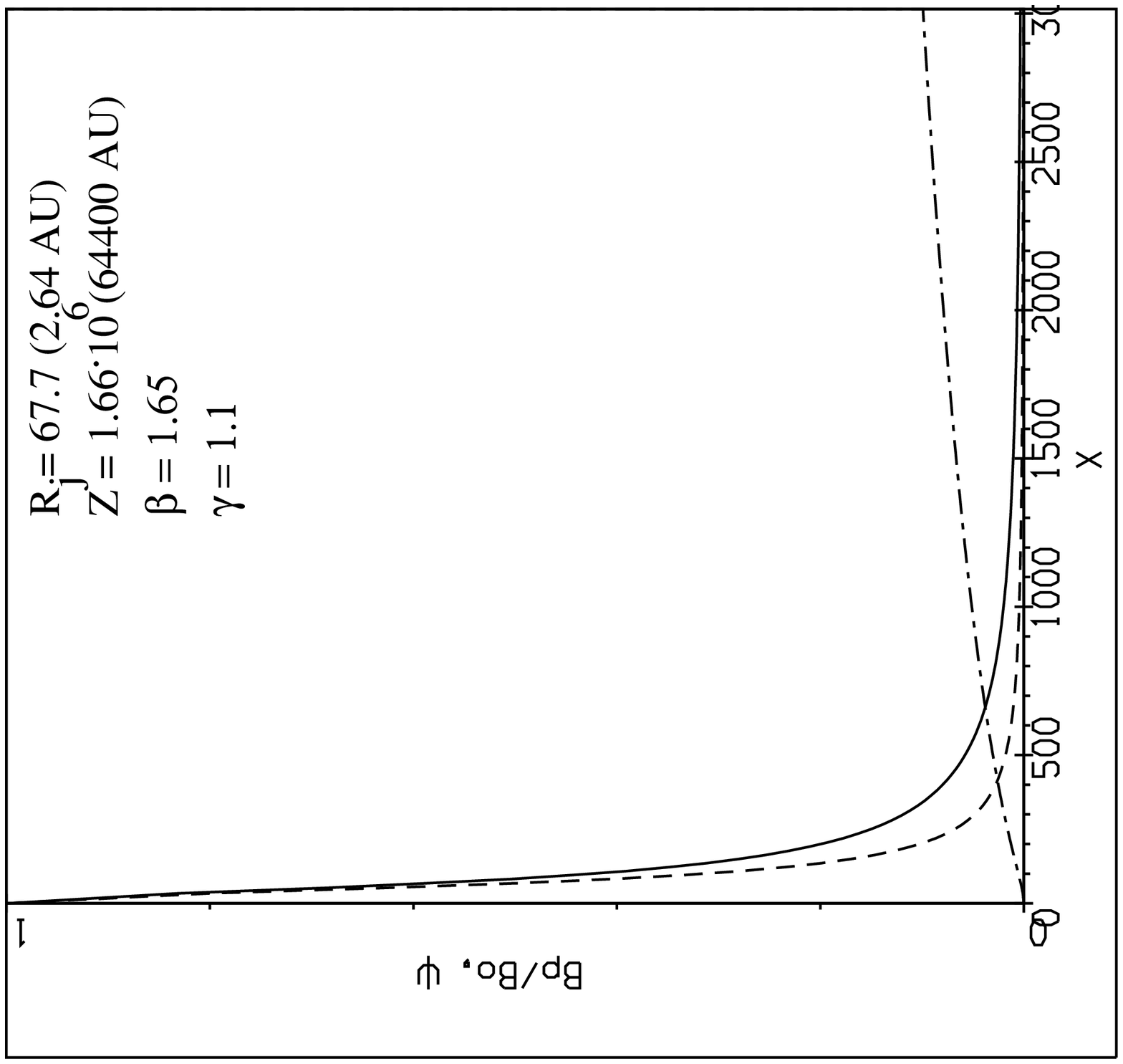,height=7.0truecm,angle=270}}

\section{Discussion of the results}

One of the basic results from the numerical simulations reported in this 
paper as well as in Paper I is that plasma ejected by rotating 
magnetized objects has the property of self-collimation. 
No other special conditions are needed to get a collimated outflow. 
Therefore, jets should be a rather common phenomenon in astrophysics.
This is the most important prediction of the theory of magnetic 
collimation which is in pretty good agreement with observations, since  
numerous jets are observed to be associated with astrophysical objects 
of different nature.

However, a direct application of the theory of magnetic collimation 
to an isotropic solar wind predicts an increase of the mass efflux near 
the poles while observations 
of the distribution of the solar wind mass efflux with heliolatitude at 
distances 5 - 7 AU by a range of instruments have given the opposite result.
In this work we eliminated this apparent contradiction by taking into 
account in the calculations more realistic parameters for the anisotropic 
solar wind, 
as those inferred recently by ULYSSES and SOHO. Then, a recalculation of the
heliolatitudinal distribution of the SW mass efflux at large distances 
with an initially anisotropic distribution of the density and flow speed 
at the base show that the initial excess of the mass flux near the equator
is preserved up to the outer boundary of the SW. 
Magnetic collimation is remarkable only in the region of the lower speed
wind near the equator. It results in a relative decrease of the mass efflux 
in comparison with the flux at the low latitudes less than $15^{o}$.
Can this effect be found in the SWAN data? To answer this question 
a detailed comparison of the calculations with these data is necessary.

It is widely believed that due to the observed close disk-jet connection 
collimated outflows are possible only from systems containing accretion 
disks. It is clearly demonstrated in this study that fast rotating stars 
without disks can also produce jet-like outflows.
This prediction is crucially important for the theory of magnetic 
collimation, since it gives a most reliable observational test of 
the theory. Up to now there has been no direct observational evidence 
that observed jets are really collimated by the magnetic field.
And, observation of jet-like outflows from objects which do not contain 
accretion disks would be this needed evidence. 

Young fast rotating stars of solar mass are among the objects which 
produce jets. 
The effect of magnetic self-collimation of the winds is parameterized 
by $\alpha$, which can be presented in the form
\begin{equation}
\alpha = 0.12 {(\psi_t/\psi_{\odot})(\Omega/\Omega_{\odot})\over 
\sqrt{(\dot M/\dot M_{\odot})}
(V_{\infty}/V_{\odot})^{3/2}}\,,
\end{equation}
for $V_{\odot}\approx  400 ~$ km/s. 
Some young stars with solar mass have angular velocities up to 100 times 
the angular velocity of the Sun (Bouvier et al. 1997).  
For example, in AB Doradus 
$\Omega_{\star} = 54~\Omega_{\odot}$ (Jardine et al. 1999). 
A phenomenological dynamo mechanism (MacGregor 1996) predicts a magnetic 
flux which scales as 
$\psi_{\star}/\psi_{\odot} 
\propto \Omega_{\star}/\Omega_{\odot}$.
However, the dependence of the mass flux on $\Omega_{\star}$ is not known. 
The theoretical analysis in the Weber \& Davis (1967) approximation 
shows that the mass flux is practically independent of the 
angular velocity. In such a case AB Doradus would have $\alpha \approx 
350$. On the other hand, if the mass loss rate is proportional to the 
magnetic pressure, 
$\dot M_{\star}/\dot M_{\odot} \propto B^2_{\star}/B^2_{\odot}$, 
a value of $\alpha$ which is 54 times smaller results, i.e., 
$\alpha \approx 6.5$.   
This star is certainly a rapid rotator and should produce a 
collimated outflow. However, can such jets be observed ? 
Unfortunately their mass flux may be too small and thus it may be practically 
impossible to observe these jets directly. 

Outflows from stars with much higher mass loss rates could be modelled 
with this study.  For example, B/Be stars have massive radiation driven 
winds. 
Usually these stars also rotate rapidly, at least Be stars.
Their mass loss rate lies in the range $\dot M_{\star} = 
(10^{-6}-10^{-8})M_{\odot}/$year, while their angular velocity is 
about $\Omega_{\star} \approx 20 \Omega_{\odot}$. 
The average magnetic field on the surface of B/Be stars can vary from 200 G 
to 1600 G (MacGregor 1996). Assuming that $R_{\star} = 10 R_{\odot}$, 
the magnetic flux from such a star is 
$\psi_\star = (7\times 10^3 - 5.6\times 10^{4}) \psi_{\odot}$. 
With wind speeds of the order of $V_{\infty}= 1000~~ km/s$, the  
parameter $\alpha$ lies in the range $\alpha = 0.05 - 5$.
This means that some of these B/Be stars (but not all) could be fast 
magnetic rotators producing therefore jet-like outflows. 
Due to their huge mass loss comparable to the mass loss of classical T 
Tauri stars, these jets could be more easily observable.
It is interesting in this connection to note observations by Marti et al.  
(1993) of jets from a B star in the HH 80/81 complex.

The most strikingly unwanted result obtained in this study is that 
the part of the total magnetic flux (and mass flux correspondingly) 
going in to the jet is only of the order of $1\%$, as it may be 
seen in Fig. \ref{radius}. Almost all magnetic flux goes in to 
the radially expanding wind with mass loss $\dot M_w$.
In this case the results of our calculations can not be directly 
applied to jets from YSO because in most of them it appears that the 
mass flux in the jets $\dot M_j$ is about $1\%$ of the accretion rate 
$\dot M_a$ (Hartigan et al. 1995).  If we assume that only about $1\%$ of 
the outflowing wind goes in to the jet, like in our results, then we 
will have the uncomfortably high ratio $\dot M_w /\dot M_a \sim 1$, which 
apparently should not be so high for outflows from accretion disks 
(Pelletier \& Pudritz, 1992). 
This means that some of our assumptions are not valid in the central 
source of jets from YSO. Modifications of the input parameters of the model
which provide collimation of an arbitrary high fraction of the wind 
into the jet will be discussed in our next paper.  

Finally we would like to emphasize once more the situation concerning wind 
flows from sources with $\alpha \gg 1$. It seems that the majority of 
interesting sources of outflows, such as rapidly rotating stars and 
systems with accretion disks are just those which satisfy this condition. 
Jets from such sources should be nonstationary and turbulent with 
properties which may strongly differ from the properties of the laminar 
jets conidered in this paper. The physics of such jets still remains 
to be studied.

\begin{acknowledgements}

SB was partially supported by Russian Ministry of education 
in the framework of the programm "Universities of Russia - basic research",
project N 897. This research has been supported in part by a NATO 
collaborative research grant CRG.CRGP.972857.

\end{acknowledgements}
\appendix

\section{Appendix}

We show that the dynamics of plasma in ideal MHD flows is invariant 
in relation to a reversal of the direction of the magnetic field lines 
in an arbitrary flux tube.
Firstly this property of ideal MHD flows was used for the solution of the
problem of plasma outlow from oblique rotators in pulsar conditions 
(Bogovalov 1999).
Here we simply show that it is also valid for flows in a gravitational 
field with thermal pressure.

The plasma flow in the nonrelativistic limit is described by the 
familiar set of the ideal MHD equations
\begin{equation}
\rho{\partial {\bf V}\over \partial t}+\rho ({\bf V \cdot \nabla}) {\bf V} 
=-\nabla P- \rho\nabla\Phi+{1\over 4\pi} ({\nabla}\times 
{\bf B})\times {\bf B},
\label{b1} 
\end{equation}
\begin{equation}
{\partial {\bf B}\over \partial t} = \nabla \times ({\bf V}\times {\bf B}),
\label{b2}
\end{equation}
\begin{equation} 
\nabla \cdot {\bf B} =0,
\label{b3}
\end{equation}
\begin{equation}
{\partial \rho\over \partial t}+ \nabla \cdot (\rho {\bf V}) =0,
\label{b4}
\end{equation}
where $\Phi$ is the gravitational potential and $P$ is the pressure.

Let us assume that we have some solution which is described by the
functions ${\bf B}(r, t)$, $\rho (r, t)$, ${\bf V}(r, t)$ and 
$P (r, t)$. We show that
the change of the direction of the magnetic field  in an arbitrary 
magnetic flux tube does not  change the dynamics of the plasma. 

Let us introduce a scalar function $\eta(r, t)$ with the  
property that 
$\eta=1$ everywhere except inside the choosen flux tube where 
$\eta=-1$. This function satisfies the following 2 conditions

\begin{equation}
{\bf B} \cdot \nabla \eta =0
\,,
\label{b5}
 \end{equation}
and 
\begin{equation}
{\partial\eta\over \partial t} +{\bf V}\cdot \nabla \eta=0
\,.
\label{b6}
\end{equation}
The second equation is the consequence of the frozen-in condition such 
that the value of $\eta$ is advected together with the plasma. 

Then the solution $\eta {\bf B}(r,t)$, $\rho (r, t)$, ${\bf V}(r,t)$ 
and $P(r, t)$ also satisfies the system of Eqs. (\ref{b1}-\ref{b4}). 
Indeed, the Lorentz force in the right hand side of (\ref{b1}) 
is,
\begin{eqnarray}
& & [\nabla \times (\eta {\bf B})]\times (\eta {\bf B}) = \eta [
\nabla \eta \times {\bf B} + \eta \nabla  \times {\bf B} ] \times {\bf B} = 
\nonumber \\ & & 
 \nabla (\eta^2/2) \times {\bf B} + \eta^2 (\nabla  \times {\bf B} ) 
\times {\bf B} = (\nabla \times {\bf B}) \times {\bf B} 
\,,
\end{eqnarray}
since $\eta^2=1$. 
%does not contain $\eta$. The right
%hand side of this equation contains this function in the combination
%$B_i \eta {\partial B_k \eta\over \partial x_l}$ 
%which can be transformed as follows
%\begin{equation}
%B_i \eta {\partial B_k \eta\over\partial x_l}=B_i \eta^2 
%{\partial B_k \over\partial x_l}+B_i B_k {1\over 2} {\partial \eta^2\over
%\partial x_l}=B_i {\partial B_k \over\partial x_l}
%\end{equation}
This means that the forces affecting the plasma do not change with this
transformation. 
Eqs.  (\ref{b2}, \ref{b3}) are satisfied due to conditions
(\ref{b5}, ~\ref{b6}). 

%\end{document}
%\newpage

\end{document}